\documentclass{ieeeaccess}
\usepackage{xcolor}
\usepackage{cite}
\usepackage{amsmath,amssymb,amsfonts}
\usepackage{algorithm}
\usepackage{algorithmic}
\usepackage{textcomp}
\usepackage{float}
\usepackage{soul}
\usepackage{changepage}
\usepackage{graphicx}
\usepackage{subcaption}
\usepackage{multirow}
\usepackage{comment}
\usepackage{threeparttable}
\usepackage{array} 

\definecolor{blue}{RGB}{0,0,255}
\usepackage{CJKutf8}
\usepackage{hyperref} % Load hyperref after all other packages
\hypersetup{
  colorlinks=false,
  linkbordercolor=white,
  urlbordercolor=white,
  pdfborder={0 0 0}
}
\def\BibTeX{{\rm B\kern-.05em{\sc i\kern-.025em b}\kern-.08em
T\kern-.1667em\lower.7ex\hbox{E}\kern-.125emX}}

\usepackage{pict2e}
\newsavebox{\ORCIDlogo}
\savebox{\ORCIDlogo}{%
\setlength{\unitlength}{\dimexpr 1em/256\relax}%
\begin{picture}(256,256)%
  \color[HTML]{A6CE39}\put(128,128){\circle*{256}}%
  \color{white}%
  \put(78.6,199.2){\circle*{20}}%
  \moveto(70.9,176.9)\lineto(86.3,176.9)\lineto(86.3,69.8)\lineto(70.9,69.8)%
  \closepath\fillpath%
  \moveto(108.9,176.9)\lineto(150.5,176.9)%
  \curveto(190.1,176.9)(207.5,148.6)(207.5,123.3)%
  \curveto(207.5,95.8)(186,69.7)(150.7,69.7)%
  \lineto(108.9,69.7)%
  \closepath\fillpath%
  \color[HTML]{A6CE39}%
  \moveto(124.3,83.6)\lineto(148.8,83.6)%
  \curveto(183.7,83.6)(191.7,110.1)(191.7,123.3)%
  \curveto(191.7,144.8)(178,163)(148,163)%
  \lineto(124.3,163)%
  \closepath\fillpath%
\end{picture}%
}
\newcommand\orcidicon[1]{\href{https://orcid.org/#1}{\usebox{\ORCIDlogo}}}

\begin{document}
\history{Date of publication xxxx 00, 0000, date of current version xxxx 00, 0000.}
\doi{10.1109/ACCESS.2017.DOI}
%\title{Machine Learning Based Parkinson Disease Detection Using in-air Dynamics Features Extraction and Selection}

%\title{Parkinson Disease Detection via In-Air Handwriting Dynamics Feature Extraction and Selection with Machine Learning}

\title{Parkinson Disease Detection Based on In-air Dynamics Feature Extraction and Selection Using Machine Learning}

%\title{Parkinson Disease Detection Using in-air Dynamics Feature Extraction and Selection Based on Machine Learning}
%\title{Parkinson's Disease Detection Using Machine Learning on Handwriting in-air Dynamics Feature Extraction and Selection}
%\title{Parkinson Disease Detection Based on in-air Dynamics Feature Extraction and Selection Using Machine Learning}

%ith Machine Learning Approaches
%\title{Effectiv feature extraction and classification approach for computer-aided Parkinson disease detection}
\author{\uppercase{Jungpil Shin}\authorrefmark{1}, \IEEEmembership{Senior Member, IEEE},
\uppercase{Abu Saleh Musa Miah \authorrefmark{1},\IEEEmembership{Member, IEEE}}, \uppercase{Koki Hirooka \authorrefmark{1},\IEEEmembership{Graduate Student Member, IEEE}},
\uppercase{Md. Al Mehedi Hasan \authorrefmark{2},\IEEEmembership{Member, IEEE}},
\uppercase{Md. Maniruzzaman} \authorrefmark{3}
}
\address[1]{School of Computer Science and Engineering, The University of Aizu, Aizuwakamatsu, Japan (e-mail:musa@u-aizu.ac.jp}
\address[2]{Department of Computer Science \& Engineering, Rajshahi University of Engineering \& Technology, Rajshahi-6204, Bangladesh (e-mail: mehedi\_ru@yahoo.com)
}
\address[3]{Statistics Discipline, Khulna University, Khulna-9208, Bangladesh, (e-mail: monir.stat91@gmail.com)}

\tfootnote{This work was supported by the Competitive Research Fund of The University of Aizu, Japan.}

\markboth
{....}
{This paper is currently under review for possible publication in IEEE Access.}

\corresp{Corresponding author: Jungpil Shin (e-mail: jpshin@u-aizu.ac.jp)}

\begin{abstract}

Parkinson's disease (PD) is a progressive neurological disorder that impairs movement control, leading to symptoms such as tremors, stiffness, and bradykinesia. Early and accurate PD detection is essential for effective management and improving patient outcomes. Many researchers analyzing handwriting data for PD detection typically rely on computing statistical features over the entirety of the handwriting task. While this method can capture broad patterns, it has several limitations, including a lack of focus on dynamic change, oversimplified feature representation, lack of directional information, and missing micro-movements or subtle variations. Consequently, these systems face challenges in achieving good performance accuracy, robustness, and sensitivity. To overcome this problem, we proposed an optimized PD detection methodology that incorporates newly developed dynamic kinematic features and machine learning (ML)-based techniques to capture movement dynamics during handwriting tasks. In the procedure, we first extracted 65 newly developed kinematic features from the first and last 10\% phases of the handwriting task rather than using the entire task. This novel approach helps reduce complexity while focusing on the phases that often exhibit significant variations in acceleration, deceleration, and directional changes—subtle movements that traditional methods may struggle to detect. Alongside this, we also reused 23 existing kinematic features, resulting in a comprehensive new feature set. Next, we enhanced the kinematic features by applying statistical formulas to compute hierarchical features from the handwriting data. This approach allows us to capture subtle movement variations that distinguish PD patients from healthy controls. To further optimize the feature set, we applied the Sequential Forward Floating Selection method to select the most relevant features, reducing dimensionality and computational complexity. Finally, we employed a ML-based approach based on ensemble voting across top-performing tasks, achieving an impressive 96.99\% accuracy on task-wise classification and 99.98\% accuracy on task ensembles, surpassing the existing state-of-the-art model by 2\% for the PaHaW dataset. This exceptional accuracy underscores the transformative potential of our approach in redefining the benchmarks for PD detection. 
\end{abstract}


\titlepgskip=-15pt

\maketitle

\section{Introduction}
\label{sec:introduction}
\PARstart{P}{arkinson's disease(PD)}  is one of the most common neurodegenerative disorders \cite{intro-1}. PD affected approximately 6 million people worldwide in 2016 \cite{intro-2}, and prevalence rates are expected to increase further as the population ages. PD is most common in people above 60 years old; on the other hand, PD can develop before the age of 40 years old (juvenile PD). Some of those include cases that may have developed due to genetic factors. The development of PD is caused by the decline in the production of a chemical that helps the brain communicate with other parts of the body. This chemical, dopamine, is also responsible for the control and fluency of movements. As the condition worsens, many brain regions are impacted. It is well known that some PD non-motor symptoms can develop years before the disease manifests. In addition, there is currently no complete cure for PD, and the main approaches are symptomatic treatments aimed at slowing the progression of the disease and alleviating symptoms. This means that diagnosing this condition at an early stage is very important to prevent its progression. Generally, the severity of PD is evaluated using the Unified Parkinson's Disease Rating Scale (URDRS). However, there is no objective quantitative method of clinical diagnosis\cite{intro-3}, and it is assessed visually and by feel by the diagnostician. Therefore, even movement disorder specialists have shown that approximately 25 \% of PD patients are misdiagnosed\cite{intro-4}.

Based on the above, developing an expert system for the analysis and diagnosis of PD is required. There are four main symptoms of PD: Tremor, which occurs when limbs are at rest; rigidity to the passive stretch of muscles; Akinesia (or bradykinesia) is a slowness in the movement or initiation of movement; and Postural instability due to loss of postural reflexes anad their symptoms include movement disorders such as hypotension, dysphagia, and freezing, and non-motor disorders such as autonomic nervous system disorders, dementia, and sleep disorders which are important diagnostic clues. Various methods have been proposed so far, such as a system using motion biometrics related to walking function\cite{intro-5}. It is well known that handwriting problems are related to this disease as well as to its severity, so changes in writing can be considered a prominent Kinematic features \cite{intro-6,intro-7}. In fact, PD \cite{related-1,related-2,related-3,related-4} and other disease\cite{intro-AD1,intro-AD2,miah2021alzheimer,hassan2024residual_miah_alzh,kafi2022lite_kidney_miah,haque2024multi_heart_disease}, a similar neurodegenerative disease, affect the structure and function of specific brain regions. It is conceivable that handwriting tasks involving kinesthetic and perceptual movements may be related to disease and its severity\cite{intro-7}. For example, the handwriting of PD patients is often characterized by reduced letter size due to micrographia \cite{intro-9} caused by finger tremors. Others included Changes in kinematic aspects of movement such as reduced speed and acceleration, increased number of speed changes, and increased travel time\cite{intro-8}.

There are two approaches to recognition: 1) online and 2) offline handwriting. Recently, deep learning (DL) approaches are often adopted, regardless of whether they are online or offline. Diaz et al. \cite{intro-13} proposed the Sequence-based dynamic handwriting analysis model consisted of 1D convolutions and BiGRUs and reported a higher performance compared to state-of-the-art at the time. However, it is often extremely difficult to gather enough data to train DL-based models. On the other hand, offline handwritten character analysis using convolutional neural networks has made it possible to analyze neural networks for small datasets by using a transfer learning approach, even if there is little data. 
%Gazda et al. \cite{intro-11} tuned the pre-trained CNN architecture using a large handwritten character dataset (MNIST, UJIpenchars2) to optimize the CNN weights for the classification of handwritten character analysis, and furthermore tuning the top layers of CNN using the target dataset. They achieved 94.7\% classification accuracy in this way.
%Afroz et al. \cite{intro-12} also experimented using CNN transfer learning. They used the five CNNs(VGG16/19, Xception, DenseNet121, MobileNetV2) and report the 94.93\% classification accuracy using VGG16. Then, they proposed the ensemble method which is a combination of VGG16 and another CNN. finally, They reported the 95.5\% classification accuracy using VGG16+Xception. 
Drotár et al. \cite{related-1} proposed a handwriting-based PD recognition system, where they extracted 13 features, including trajectory, velocity, jerk, and stroke. By applying a Support Vector Machine (SVM) with an radial basis function (RBF) kernel, they achieved a classification accuracy of 79.4\% in distinguishing PD patients from healthy controls. In a subsequent study, Drotár et al. \cite{related-2} expanded their feature sets to 600 features while maintaining the same eight combined tasks. They implemented a Whitney U-test filter and a relief algorithm for feature selection, resulting in an improved classification accuracy of 80.09\%. Drotár et al. \cite{related-3} further advanced their approach by utilizing seven tasks and employing entropy, signal energy, and empirical mode decomposition (EMD) as feature types. With the Mann-Whitney U-test and relief algorithm for feature selection, they reported an impressive accuracy of 88.1\%.

Another study, Drotár et al. \cite{related-7} analyzed seven tasks using entropy and EMD features without any feature selection algorithm, achieving an area under the curve (AUC) of 89.09\%. Similarly, Mucha et al. \cite{related-5} conducted an extensive analysis involving eight tasks, where they included velocity, acceleration, frequency domain features, and fractional-order derivatives. Their model utilized XGBoost for classification, attaining a remarkable accuracy of 97.14\%. Impedovo et al. \cite{related-4} also contributed significantly to this field, analyzing eight tasks with a baseline of 24 features and using an SVM-linear classifier with 10-fold cross-validation (CV), which yielded an accuracy of 88.33\%. They further refined their methodology by incorporating new features, yielded a higher accuracy of 93.79\%. In a focused approach on three tasks, they combined 24 baseline features with 3 new features, achieving an accuracy of 97.14\%.

The main problem of the existing feature extraction study mentioned above \cite{related-1,related-2,related-3,related-4,related-5,related-7} is that they often rely on the entire phase of the task, and some phases of the task of ten exhibit significant variations in acceleration, deceleration, and directional changes—subtle movements that are difficult to detect using the mentioned methods. Although the entire handwriting task phase used in the existing study to extract the features may capture broad patterns. However, they did not mention the dynamic change of handwriting movement \cite{related-1,related-2,related-3,related-4,related-5,related-7}. Moreover, there is a lack of focus on dynamic changes, oversimplified feature representation, insufficient directional information, and neglect of micro-movements or subtle variations. Consequently, existing systems struggle with performance accuracy, robustness, and sensitivity due to lacking effective features. 

To address these challenges, we proposed an optimized PD detection methodology that incorporates newly developed dynamic kinematic features and machine learning techniques to capture movement dynamics during handwriting tasks. We the newly developed movement dynamic 65 kinematic feature to capture the subtle movement changes occurring during the initial and final stages of writing, which may differ significantly between PD patients and healthy controls (HC). Our main contributions to this study are given below:

\begin{itemize}
    \item \textbf{New Dynamic Feature Extraction Method:} 
    
    We extracted 65 new kinematic features from the first and last 10\% phases of the handwriting task, rather than using the entire task, to reduce complexity. These phases are crucial as they often show significant variations in acceleration, deceleration, and directional changes—subtle movements that conventional methods may struggle to detect. By isolating these segments, our approach can capture critical transitions and irregularities in movement that are more pronounced in PD patients. These features include angle trajectory, signed x/y displacement, velocity, first/last displacement, etc. Moreover, we introduce angle features related to handwriting and a directional speed feature, which accounts for the movement direction rather than relying solely on the absolute speed values in the x- and y-axes.  Alongside these 65 newly introduced features, we also reused 23 existing kinematic features, resulting in a comprehensive feature set designed to improve PD detection accuracy. As we are using the first and last phases of the handwriting tasks it makes completely new features for this also. We extracted these features to address the limitations in existing feature sets for detecting PD-related movement characteristics across multiple tasks.

\item \textbf{Hierarchical Feature Extension } 
 We enhanced the kinematic features by applying statistical theorems, including central tendency, dispersion, and higher-order relationships, to compute hierarchical features from the handwriting data. This hierarchical feature leads to gaining deeper insights into the writing process from each Kinematic feature. By integrating the newly extracted features with the existing ones, we calculated around $(65+23) \times (11 \pm 3) = 844$ comprehensive and effective feature set that significantly improves the detection of PD-specific movement patterns during handwriting tasks.

\item \textbf{Feature Selection with Sequential Forward Floating Selection (SFFS):}

We employ SFFS to identify the most impactful robust features, refining the feature set to ensure optimal performance across different ML-based classifiers. This selection process enhances the system's ability to focus on the most relevant features for PD detection.

\item \textbf{Classifier Optimization with Optuna and Ensemble Voting:} We employed leave-one-out-cross validation (LOOCV) for splitting the PaHaW dataset into training and testing. By leveraging Optuna for classifier optimization, we fine-tuned ML-based models for PD detection. Furthermore, ensemble voting across top-performing tasks increased the robustness and accuracy of our system, achieving an outstanding performance accuracy of  96.99\%  and 99.98\% accuracy for individual task accuracy average and tasks ensemble, respectively  for the PaHaW dataset. This demonstrates the effectiveness and reliability of our approach, setting a new benchmark in PD detection methodologies. 
\end{itemize}

The paper is organized as follows. After the introductory section, the description of the used database is given. Next, the methodology of extracting the features from handwriting signals is given, followed by a brief overview of the used classifier. Finally, the numerical results and conclusions are provided. Our contributions to the study are given below:

\section{Related Works}

PD/HC classification problem by handwriting has long been a hot topic of research, and various features have been proposed to identify the impact of PD on handwriting. Drotár et al.\cite{related-1} investigated features related to speed (stroke speed, velocity, acceleration, jerk, etc.) and the average number of local areas during task and writing (NCV, NCA). They calculated the Pearson correlation coefficients between feature vectors and associated responses and showed that there was a strong correlation between certain feature vectors and responses, such as stroke speed and stroke width.

Also, Drotár et al.\cite{related-2} tried to apply not only the information on-surface but also the trajectory in the air during the task to predict PD. Similarly, when the correlation coefficient between each feature vector and the response was calculated, eight feature vectors out of the top 10 feature vectors were feature vectors related to in-air trajectories during writing, indicating that there was a strong correlation. 

In the paper, \cite{related-3}, entropy, energy, and empirical mode decomposition were proposed as new feature vectors in addition to the existing feature vectors. These feature vectors showed a higher absolute correlation coefficient than the existing feature vectors, and using the feature vectors selected by the Relief algorithm, we achieved high classification accuracy with the RBF kernel SVM.

Moetesum et al.\cite{related-6} proposed a novel method of assessing their contribution to the characterization of PD. Hypothetically, they thought about the importance of visual features. They used convolutional neural networks to extract discriminating visual features. Convolutional Neural Networks are employed to extract discriminating visual features. Classification is carried out using SVM model and predictions of different tasks are combined using majority voting (late fusion). Evaluations on a standard dataset of 72 subjects reported an overall a classification accuracy of 83\%.

In the paper \cite{related-5}, they proposed fractional-order derivative (FD) handwriting features for investigating the relationship with patient's clinical data. The utilization of the FD as a substitution for the conventional differential derivative during the calculation of the basic kinematic features provides a new advanced approach. In comparison with the conventional kinematic features, They showed FD-based ones correlate more significantly with the clinical characteristics (UPDRS V and PD duration). Especially, They showed strong correlations for handwriting tasks based on the periodic repetition of specific movements (Archimedean spiral; repetitive letter l, syllable le, or word les). In addition, when they used XGBoost to classify binary classification tasks of HC and PD, they achieved the highest classification accuracy (97.74\%).

Impedovo\cite{related-4} also adopted new capabilities for classifying PD using functions such as the sigma-lognormal model, Maxwell-Boltzmann distribution, and discrete Fourier transform. Those features were also used to classify the same neurodegenerative disease, Alzheimer's disease.  The classifier used in the experiment was an SVM with a linear kernel, and its generalization performance was confirmed by 10-fold CV. As a result, Sigma-Lognormal and Maxwell-Boltzmann feature vectors were included in the 10 most relevant, and achieved the classification accuracy  of 98.44\% by combining the top 3 tasks with the highest classification accuracy.

Parziale et al. \cite{related-8} proposed the method adopting the Negative Selection Algorithm(NSA). Their approach has mainly two advantages compared with previous approach. One is the training. This model requires only healthy control data, thus avoiding the burden of collecting patients' data. They reported the average classification accuracy of 97.12\%.

Krut et al. \cite{related-9} focused on the Archimedean spiral drawings task and adopted Dynamic Time Warping (DTW) to match optimally between individual data and reference spiral and to calculate the Euclidean distance values between them. Then, they applied SVM and k-nearest neighbors (k-NN) to discriminate patients PD from HC. They reported that the highest the classification accuracy of 97.52\% was obtained by SVM model.

Deharab et al. \cite{related-10} also adopted a nonlinear method based on Dynamic Writing Traces Warping (DWTW) in combination with a SVM model. They measured the performance using the PaHaW dataset and reported 88.33\% accuracy for task 8 among PaHaW's tasks. They finally concluded it offered a superior trade-off between correct diagnosis and computational difficulty compared with previous approaches, making it a guaranteed method for clinical usage.

The main problem of the existing feature extraction study mentioned above \cite{related-1,related-2,related-3,related-4,related-5,related-7} is that they often rely on statistical features computed over the entirety of the handwriting task. This entire handwriting task used for the feature extraction approach may capture broad patterns. However, it has significant limitations, including a lack of focus on dynamic changes, oversimplified feature representation, insufficient directional information, and neglect of micro-movements or subtle variations. Consequently, existing systems struggle with performance accuracy, robustness, and sensitivity. To address these challenges, we proposed an optimized methodology for PD detection by taking the first 10\% and last 10\% handwriting of each task alongside kurtosis and skewness indices to overcome the challenges. In addition, we leverages newly developed movement dynamic features to capture the subtle movement changes occurring during the initial and final stages of writing, which may differ significantly between PD patients and HC.

\section{Materials and Methods}

\subsection{Proposed Methodology}
\begin{figure*}[htbp]
\begin{adjustwidth}{-1cm}{0cm}
        \centering 
    \includegraphics[width=0.90\linewidth]{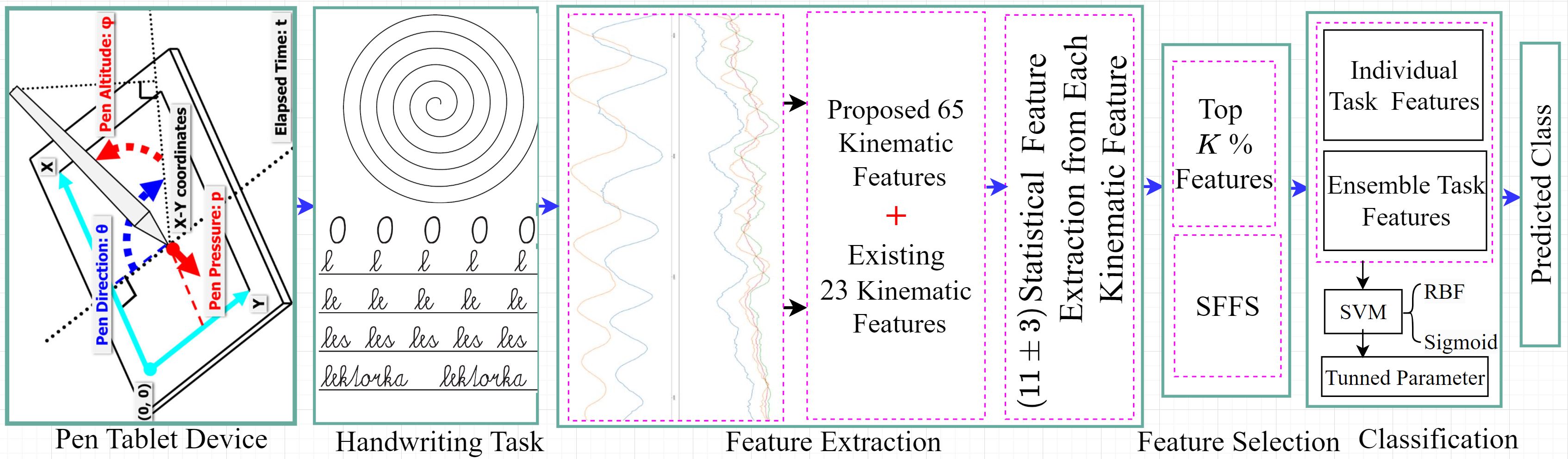}
    \caption{Our proposed model architecture for PD detection.}
    \label{fig:Our_Experiment_Flow}
    \end{adjustwidth}
\end{figure*}

The overall system workflow architecture is illustrated in Fig. \ref{fig:Our_Experiment_Flow}. Despite significant efforts by researchers to improve performance accuracy using the benchmark PaHaW handwriting dataset, challenges remain in achieving optimal results \cite{related-1,related-2,related-3,related-4,related-5,related-7}. These issues arise from a lack of focus on dynamic changes, oversimplified feature representation, insufficient directional information, and neglect of micro-movements or subtle variations. As a result, existing systems struggle with performance accuracy, robustness, and sensitivity. To address these challenges, we propose an optimized methodology for PD detection, incorporating newly developed dynamic kinematic features and machine learning techniques to capture the movement dynamics during handwriting tasks. The system consists of three main stages:

\begin{itemize}
    \item 
\textbf{Stage 1: Feature Extraction} \\
In this stage, we newly proposed 65 dynamic kinematic features extraction approach from the first and last 10\% phases of the handwriting task, instead of using the entire task. These phases are crucial because they exhibit significant variations in acceleration, deceleration, and directional changes—subtle movements that conventional methods often fail to detect. By isolating these segments, our approach captures critical transitions and movement irregularities that are more pronounced in PD patients. The extracted features include angle trajectory, signed x/y displacement, velocity, first/last displacement, and others. We also introduce new angle and directional speed features, which capture movement direction rather than relying solely on absolute speed values in the x- and y-axes. Additionally, we incorporated 23 existing kinematic feature extraction methods, enhanced by focusing on the initial and final phases of the handwriting task, providing a completely new set of features.
\item 
\textbf{Stage 2: Hierarchical Feature Extension}\\
We extended the kinematic features by applying statistical theorems, including central tendency, dispersion, and higher-order relationships, to compute hierarchical features from the handwriting data. This enables deeper insights into the writing process, enhancing our understanding of PD-specific movement patterns. The resulting comprehensive feature set, consisting of both newly developed and existing features, is augmented with approximately 844 statistical features derived from the original 88 kinematic features.

\textbf{Stage 3: Feature Selection and Model Optimization} \\

To reduce computational complexity and improve performance, we applied Sequential Forward Floating Selection (SFFS) to identify the most informative features, minimizing redundancy and ensuring optimal performance across different classifiers. Additionally, SVM hyperparameters were optimized using Optuna to enhance classifier robustness. Finally, ensemble voting was performed across top-performing tasks, including top-3, top-5, and all tasks, to evaluate and maximize system performance.
\end{itemize}

This approach ensures the system achieves high accuracy, sensitivity, and robustness in detecting PD, outperforming existing state-of-the-art methods. The details description of the each procedure are given below.

% \Figure[hbtp]{scale=0.05}{Figures/PDHandWriting_System_flow.png}{PDHandWriting System Flow}
%In this study, we aim to enhance traditional handwriting analysis approaches for Parkinson’s Disease (PD) detection by introducing novel statistical features that capture subtle and nuanced aspects of movement dynamics.

\subsection{PaHaW Dataset}

Experimental evaluation has been performed on the PaHaW dataset\cite{PaHaW1}. It includes 37 PD patients and 38 HC. Tasks include words written in Czech (the participants' native language). Each participant completed 8 handwriting tasks: 1. drawing the Archimedes spiral; 2. writing in cursive the letter “l”, 3. the bigram “le” and 4. the trigram “les”; 5. writing in cursive the words “lektorka” (female teacher in Czech), 6. “porovnat” (to compare), and 7. “nepopadnout” (to not catch); 8. writing in cursive the sentence “Tramvaj dnes uz nepojede” (The tram won’t go today). A digitizing tablet (Wacom Intuos 4M) was overlaid with an empty paper template, and participants were allowed to repeat a task if they made mistakes. Online handwriting signals were recorded with fs = 150 Hz sampling rate.  The raw data captured by the device are the x- and y- y-coordinates of the pen position and their timestamps. Moreover, measures of pen inclination, i.e., azimuth and altitude, and the pressure were recorded. The last signal concerns the so-called button status, which is a binary variable evaluating 0 for the pen-up state (in-air movement) and 1 for the pen-down state (on-surface movement). In this context, this dataset collection method adopted a Pen-Tablet device for analyzing PD. The main advantage of online acquisition devices is their ability to acquire the kinematics (dynamics) of the writing process, which are lost in offline systems. In this case, the trait is represented as a sequence $ {S(n)}_{n = 0, 1,..., N} $, where $ S(n) $ is the signal value sampled at time $ n \Delta t $ of the writing process ($0 \leq n \leq N$), $ \Delta t $ is the sampling period.

\subsection{Feature Extraction}

In this study, we focus on enhancing the accuracy and robustness of PD detection by extracting kinematic features and then extending them with statistical formulas to calculate hierarchical features from handwriting tabular data. While existing methods have used traditional kinematic features, such as pressure, azimuth, altitude, displacement, velocity, and stroke count, these often fail to capture subtle and nuanced movement patterns. Traditional kinematic features typically consider the tabular device information and their entire range of data, calculating values such as x, y,z, azimuth, altitude, displacement or velocity over the full task duration. However, such methods overlook the fine details of handwriting motion that can vary across different phases of the writing task. This limitation may reduce sensitivity when distinguishing between PD patients and healthy controls, especially in terms of micro-movements or variations in behaviour during specific phases of the handwriting task.

To address these challenges, we introduce 65 new dynamic movement-based kinematic features in addition to the 23 baseline features from previous studies. These new features focus on capturing detailed variations in handwriting dynamics, which are critical for improving detection accuracy. Our proposed kinematic feature extraction approach specifically targets the mentioned limitations by focusing on key segments of the handwriting process. We introduce features that are computed from designated portions of the recorded data, such as the first 10\% and last 10\% of the writing sequence. These phases are particularly important as they often exhibit significant variations in acceleration, deceleration, and directional changes—subtle movements that are difficult to detect using conventional methods. By isolating these segments, our approach is able to capture critical transitions and irregularities in movement that are more pronounced in PD patients.

For example, angle trajectory, signed displacement and velocity variations track directional changes and speed fluctuations during writing, offering insights into spatial inconsistencies and motor control deficits that may differ between PD patients and healthy individuals. Additionally, first/last displacement highlights differences in starting and ending positions, which can indicate unique motor impairments seen in PD patients. These new features delve deeper into the dynamic aspects of handwriting, moving beyond general movement patterns and providing greater sensitivity to subtle variations.
Table \ref{tab:Proposed_New_Features} summarizes the newly introduced features. By incorporating these into our analysis pipeline, we aim to improve not only classification accuracy but also model interpretability and robustness. These advanced kinematic features are designed to enhance the ability to detect subtle irregularities in handwriting, leading to better performance in PD detection systems. Table \ref{tab:Baseline features} provides an overview of the features adopted by state-of-the-art approaches, with more details on their implementation found in \cite{intro-3,intro-7,related-1,related-2,related-3,related-4}. Here, we first extracted kinematic feature then we extended function-based features by evaluating statistical parameters such as mean, median, standard deviation, 1st and 99th percentiles, kurtosis, and skewness. These extensive level of the feature extraction approach lead to the robust representation of handwriting dynamics and improve the overall detection model.

\subsubsection{Our Proposed Kinematic Feature Extraction Approaches}
Table \ref{tab:Proposed_New_Features} summarizes the new kinematic features proposed in this paper, which are designed to capture subtle changes in movement, particularly during the initial and final stages of handwriting. These features are enhanced with statistical parameters, enabling a deeper understanding of how handwriting dynamics differ between PD patients and HC. Specifically, these features provide valuable insights into how PD patients start and finish their handwriting tasks, revealing distinctive patterns that may be indicative of the disease.

The newly introduced features are applied to key speed-related metrics, such as displacement and velocity, to track movement variations. By focusing on these dynamic aspects of handwriting, we aim to capture fine-grained differences that can help distinguish PD patients from healthy individuals. The new kinematic feature calculation procedure is described in detail below.

\paragraph{Angle in Trajectory}
First, we extracted the angle trajectory feature aiming to extract tremors and rigid-based information. Tremors and rigid muscles are well-known symptoms of PD. Even if the content is smooth for HC, such as straight lines and curves, PD may not be able to write smoothly due to these symptoms, and noise may be included in the trajectory. We thought we could identify it by calculating the curvature of the pen's trajectory to obtain this information.

 Intuitively, we can decide any time $t$, and before time $t-1$, and after time $t+1$.The pen position at each time can be expressed as $p_t=(x_t,y_t)$, $p_{t-1}=(x_{t-1},y_{t-1})$ and $p_{t+1}=(x_{t+1},y_{t+1})$. Then we calculate two vectors ($\vec{v_1} = p_{t-1} -p_t$, and $\vec{v_2} = p_{t+1} -p_t$) with reference to time $t$. By using these two vectors and the inner product formula, the degree of curvature (angle) at time t can be calculated. Calculate angle formula is as follows:
 $$ \cos\theta = \frac{\vec{v_1}\cdot\vec{v_2}}{\|\vec{v_1}\|\|\vec{v_2}\|}$$
 However, angle calculations at adjacent time intervals such as $t$ and $t+1$ are often susceptible to small noises. So, by calculating the angle between the point of based time $t$ and the point left any length(along the trajectory) from the point of based time $t$, we can change the magnitude of the noise affected. In this experiment, we calculated the angle from the based time $t$ (based point $p_t$) to any point $d (px/nm/mm) \in (10,20,......,100)$ in trajectory distance away.

\paragraph{Signed x / y displacement:}
This feature captures the signed displacement in the x and y directions, where the sign (+/-) retains information about the direction of movement. This distinction helps in understanding the directional aspects of the pen's movement during handwriting. The displacement formula is given by equation \ref{eq:signed_displacement}:
\begin{equation}
d_i = \frac{x_{i+1}(y_{i+1}) - x_i(y_i)}{t_{i+1} - t_i}, \;\; 1 \leq i \leq n - 1
\label{eq:signed_displacement}
\end{equation}

\paragraph{Signed x / y velocity:}
This feature calculates the signed velocity in the x and y directions, accounting for the direction of movement. It helps in analyzing the speed dynamics in both horizontal and vertical axes, offering insights into handwriting anomalies.

\paragraph{First Displacement:}
The first displacement represents the initial movement of the pen, capturing how far the pen moves in the early stages of the task. The formula is given by equation \ref{eq:first_displacement}:
\begin{equation}
d_i = \frac{\sqrt{(x_{i+1} - x_i)^2 + (y_{i+1} + y_i)^2}}{t_{i+1} - t_i}, \;\; 1 \leq i \leq n \times 0.1
\label{eq:first_displacement}
\end{equation}

\paragraph{First Velocity:}
First velocity focuses on the speed at the beginning of the handwriting task, reflecting how quickly the pen moves during the initial movements. The velocity is computed using equation \ref{eq:first_velocity}:
\begin{equation}
v_i = \frac{d_i}{t_{i+1} - t_i}, \;\; 1 \leq i \leq n \times 0.1
\label{eq:first_velocity}
\end{equation}

\paragraph{First x / y displacement:}
This feature calculates the magnitude of displacement specifically in the x and y directions, only for the first 10\% of the total movement. See equation \ref{eq:first_xy_displacement}:
\begin{equation}
d_i = \left| \frac{x_{i+1}(y_{i+1}) - x_i(y_i)}{t_{i+1} - t_i} \right|, \;\; 1 \leq i \leq n \times 0.1
\label{eq:first_xy_displacement}
\end{equation}

\paragraph{First signed x / y displacement:}
Similar to the first displacement, but with the signed value retained, offering directionality information for the initial movements in the x and y directions. The formula is given by equation \ref{eq:first_signed_xy_displacement}:
\begin{equation}
d_i = \frac{x_{i+1}(y_{i+1}) - x_i(y_i)}{t_{i+1} - t_i}, \;\; 1 \leq i \leq n \times 0.1
\label{eq:first_signed_xy_displacement}
\end{equation}

\paragraph{Last Displacement:}
This feature measures the displacement during the final stage of the handwriting task, analyzing how far the pen moves toward the end. The formula is provided in equation \ref{eq:last_displacement}:
\begin{equation}
d_i = \frac{\sqrt{(x_{i+1} - x_i)^2 + (y_{i+1} + y_i)^2}}{t_{i+1} - t_i}, \;\; n - n \times 0.1 \leq i \leq n - 1
\label{eq:last_displacement}
\end{equation}

\paragraph{Last Velocity:}
This measures the velocity during the last 10\% of the handwriting task, providing insight into the speed with which the writing is completed. The formula is given by equation \ref{eq:last_velocity}:
\begin{equation}
v_i = \frac{d_i}{t_{i+1} - t_i}, \;\; n - n \times 0.1 \leq i \leq n - 1
\label{eq:last_velocity}
\end{equation}

\paragraph{Last x / y displacement:}
This feature computes the magnitude of displacement in the x and y directions for the final part of the task, as described by equation \ref{eq:last_xy_displacement}:
\begin{equation}
d_i = \left| \frac{x_{i+1}(y_{i+1}) - x_i(y_i)}{t_{i+1} - t_i} \right|, \;\; n - n \times 0.1 \leq i \leq n - 1
\label{eq:last_xy_displacement}
\end{equation}

\paragraph{Last signed x / y displacement:}
This captures the signed displacement in the x and y directions at the end of the handwriting task, indicating how the pen moves directionally in its final stages. The displacement is given by equation \ref{eq:last_signed_xy_displacement}:
\begin{equation}
d_i = \frac{x_{i+1}(y_{i+1}) - x_i(y_i)}{t_{i+1} - t_i}, \;\; n - n \times 0.1 \leq i \leq n - 1
\label{eq:last_signed_xy_displacement}
\end{equation}

\paragraph{Last signed x / y velocity:}
This measures the signed velocity in both the x and y directions toward the end of the handwriting task, offering insights into the final speed and direction of motion.

\paragraph{Last signed x / y displacement:}
This captures the signed displacement in the x and y directions at the end of the handwriting task, indicating how the pen moves directionally in its final stages. The signed displacement formula is given by equation \ref{eq:last_signed_xy_displacement}: \begin{equation} d_i = \frac{x_{i+1}(y_{i+1}) - x_i(y_i)}{t_{i+1} - t_i}, ;; n - n \times 0.1 \leq i \leq n - 1 \label{eq:last_signed_xy_displacement} \end{equation}

\paragraph{Last x / y velocity:}
This feature calculates the velocity in the x and y directions during the last 10\% of the task, providing insights into the dynamics of pen movement at the final stage of handwriting. The formula is expressed by equation \ref{eq:last_xy_velocity}: \begin{equation} v_i = \frac{d_i}{t_{i+1} - t_i}, ;; n - n \times 0.1 \leq i \leq n - 1 \label{eq:last_xy_velocity} \end{equation}

\paragraph{Last signed x / y velocity:}
This feature computes the signed velocity in the x and y directions during the final portion of the task, emphasizing the direction and speed of pen movement. The signed velocity formula is expressed by equation \ref{eq:last_signed_xy_velocity}: \begin{equation} v_i = \frac{d_i}{t_{i+1} - t_i}, ;; n - n \times 0.1 \leq i \leq n - 1 \label{eq:last_signed_xy_velocity} \end{equation}

\paragraph{Last Pen Information:}
This group of features captures the final pen states, including the pressure, azimuth, and altitude at the end of the handwriting task. These values can provide further insights into how the pen's physical parameters behave as the writing task concludes.

\paragraph{Stroke Pressure:}
This group of features focuses on the pressure applied during each stroke. It includes several statistical parameters such as maximum, minimum, mean, median, variance, standard deviation, 1st and 99th percentiles, skewness, and kurtosis. These features provide a comprehensive view of the pressure dynamics during the handwriting task.

\paragraph{Stroke Pressure Displacement:}
This set of features focuses on the displacement of the pen during each stroke, incorporating statistical metrics such as maximum, minimum, mean, median, variance, standard deviation, 1st and 99th percentiles, skewness, and kurtosis. These features provide insights into how displacement patterns vary across individual strokes.

\paragraph{Stroke Pressure Velocity:}
This feature focuses on the velocity of the pen during each stroke, including statistical measures such as maximum, minimum, mean, median, variance, standard deviation, 1st and 99th percentiles, skewness, and kurtosis. It is designed to capture the speed variations during the strokes, offering further detail on how velocity fluctuates across different sections of the handwriting task.
% \subsubsection{Angle in trajectory}

\begin{table*}
\centering
    \caption{Proposed Dynamic Movement 65 Proposed New Features Extraction Method}
    \label{tab:Proposed_New_Features}
    \begin{tabular}{l|m{90pt}|m{300pt}} % Added an extra column for the feature sequence number
    \hline
    \textbf{No.} & \textbf{New Feature Name} & \textbf{Description} \\ % Added the header for the sequence number
    \hline \hline
             1 & Angle in Trajectory & 1-10 daa mean, 10-99 percentile, 100 median and standard deviation  \\
    \hline
    2-3 & Signed x and y displacement & In the case of Signed, the signed(+/-) can keep information on which way the pen moved. \newline The specific formula is as follows: 
    \begin{displaymath}
        d_{i} = 
        \begin{array}{l}
        \displaystyle \frac{x_{i+1}(y_{i+1}) -x_{i}(y_{i}) }{t_{i+1}-t_{i}} \;\;\; 1\leqq i \leqq n-1\\
        \end{array}
    \end{displaymath}\\

    \hline
    4-5 & Signed x and y velocity & Signed Velocity in the x / y direction. \\
        \hline
      6-8 & First Pen Information & First Pressure, First Azimuth, First Altitude,\\
    \hline
    9 & First Displacement & 
    \begin{displaymath}
        d_{i} = 
        \begin{array}{l}
        \displaystyle \frac{\sqrt[2]{( x_{i+1} - x_{i} )^{2}+( y_{i+1} + y_{i} )^{2}}} {t_{i+1}-t_{i}} \;\;\; 1\leqq i \leqq n*0.1 \\
        \end{array}
    \end{displaymath}\\
    \hline
    10 & First Velocity &  
    \begin{displaymath}
        v_{i} = 
        \begin{array}{l}\displaystyle \frac{d_{i}}{t_{i+1}-t_{i}}  \;\;\; 1\leqq i \leqq n*0.1\\
        \end{array} 
    \end{displaymath}\\ 
    \hline
    11-12 & First x and y displacement & 
    \begin{displaymath}
        d_{i} = 
        \begin{array}{l}
        \displaystyle \left| \frac{x_{i+1}(y_{i+1}) -x_{i}(y_{i}) }{t_{i+1}-t_{i}} \right| \;\;\; 1\leqq i \leqq n*0.1 \\
        \end{array}  
    \end{displaymath}\\
    \hline
    13-14 & First x and y velocity & First Velocity in the x / y direction \\
    \hline
    15-16 & First signed x and y displacement & 
    \begin{displaymath}
        d_{i} = 
        \begin{array}{l}
        \displaystyle \frac{x_{i+1}(y_{i+1}) -x_{i}(y_{i}) }{t_{i+1}-t_{i}} \;\;\; 1\leqq i \leqq n*0.1\\
        \end{array}
    \end{displaymath}\\
    \hline
    17-18 & First signed x and y velocity & First signed Velocity in the x / y direction. \\
    \hline
    19 & Last Displacement & 
    \begin{displaymath}
        d_{i} = 
        \begin{array}{l}
        \displaystyle \frac{\sqrt[2]{( x_{i+1} - x_{i} )^{2}+( y_{i+1} + y_{i} )^{2}}} {t_{i+1}-t_{i}} \;\;\; n-n*0.1\leqq i \leqq n -1 \\
        \end{array}
    \end{displaymath}\\
    \hline
    20 & Last Velocity &  
    \begin{displaymath}
        v_{i} = 
        \begin{array}{l}\displaystyle \frac{d_{i}}{t_{i+1}-t_{i}}  \;\;\; n-n*0.1\leqq i \leqq n -1 \\
        \end{array} 
    \end{displaymath}\\ 
    \hline
    21-22 & Last x and y displacement & 
    \begin{displaymath}
        d_{i} = 
        \begin{array}{l}
        \displaystyle \left| \frac{x_{i+1}(y_{i+1}) -x_{i}(y_{i}) }{t_{i+1}-t_{i}} \right| \;\;\; n-n*0.1\leqq i \leqq n -1 \\
        \end{array}  
    \end{displaymath}\\
    \hline
    23-24 & Last x and y velocity & Last Velocity in the x / y direction \\
    \hline
    25-26 & Last signed x  and y displacement & 
    \begin{displaymath}
        d_{i} = 
        \begin{array}{l}
        \displaystyle \frac{x_{i+1}(y_{i+1}) -x_{i}(y_{i}) }{t_{i+1}-t_{i}} \;\;\; n-n*0.1\leqq i \leqq n -1 \\
        \end{array}
    \end{displaymath}\\
    \hline
    27-28 & Last signed x and y velocity & Last signed Velocity in the x / y direction. \\
    \hline

        29-30 & Last x and y & Last  x / y direction. \\
    \hline
     31-33 & Last Pen Information & Last Pressure, Last Azimuth, Last Altitude,\\
    \hline

        34-44 & Stroke Pressure & Stroke Pressure X each stroke {X= max, min, mean, median, variance, standard deviation,1st, 99th, skewness, kurtosis} \\
    \hline

            44-54 & Stroke Pressure Displacement & Stroke displacement X of each stroke (X=max, min, mean, median, variance, standard deviation,1st, 99th, skewness, kurtosis) \\
    \hline
55-65 & Stroke Pressure Velocity & Stroke velocity X each stroke (X=max, min, mean, median, variance, standard deviation,1st, 99th, skewness, kurtosis) \\
    \hline
    \end{tabular}
\end{table*}

\subsubsection{Existing Kinematic Feature Extraction Approaches}
In this study, we also employed 23 existing kinematic feature extraction methods, as shown in Table \ref{tab:Baseline features}. These methods were implemented to serve as the baseline system for our analysis. 

\begin{table*}[]
    \centering
    \caption{Existing Baseline Features Extraction Method}
    \label{tab:Baseline features}
    \begin{tabular}{l|l|m{300pt}} % Added an extra column for the feature sequence number
    \hline
    \textbf{No.} & \textbf{Feature Name} & \textbf{Description} \\ % Added the header for the sequence number
    \hline
    1-2 & Position & Position in terms of s(x,y) or Coordinates\\
    \hline
    3- & Button Status & Movement in the air: b(t)=0\newline Movement on the pad: b(t)=1\\
    \hline
    4 & Pressure & Pressure of the pen on the pad\\
    \hline
    5 & Azimuth & Angle between the pen and the vertical plane on the pad\\
    \hline
    6 & Altitude & Angle between the pen and the pad plane\\
    \hline
    7 & Displacement & 
    \begin{displaymath}
        d_{i} = \left\{ 
        \begin{array}{l}
        \displaystyle \frac{\sqrt[2]{( x_{i+1} - x_{i} )^{2}+( y_{i+1} + y_{i} )^{2}}} {t_{i+1}-t_{i}} \;\;\; 1\leqq i \leqq n-1 \\
        d_{n} - d_{n-1} \;\;\; i=n
        \end{array} \right.
    \end{displaymath}\\
    \hline
    8 & Velocity &  
    \begin{displaymath}
        v_{i} = \left\{ 
        \begin{array}{l}\displaystyle \frac{d_{i}}{t_{i+1}-t_{i}}  \;\;\; 1\leqq i \leqq n-1\\
        v_{n} - v_{n-1} \;\;\; i=n
        \end{array} \right.
    \end{displaymath}\\ 
    \hline
    9-10 & x and y displacement & Displacement in the x and y direction. The specific formula is as follows.
    \begin{displaymath}
        d_{i} = \left\{ 
        \begin{array}{l}
        \displaystyle \left| \frac{x_{i+1}(y_{i+1}) -x_{i}(y_{i}) }{t_{i+1}-t_{i}} \right| \;\;\; 1\leqq i \leqq n-1 \\
        d_{n} - d_{n-1} \;\;\; i=n
        \end{array}  \right.
    \end{displaymath}\\
    \hline
    11-12 & x and y velocity & Velocity in the x and y direction \\
    \hline
 
    13 & NCV & Number of changes of Velocity \\
    \hline
    14 & NCA & Number of changes of Acceleration \\
    \hline
    15 & NCP & Number of changes of Pressure \\
    \hline
    16 & Stroke number & Number of stroke\\
    \hline
    17 & Durations & in-air time, on-surface time, total task time, Ratio of time spent in-air/on-surface\\
    \hline
    18-19 & Entropy & Shannon and Rényi operators applied on (x,y) \\
    \hline
    20-21 & Energy & Conventional Energy and Teager-Kaiser energy\\
    \hline
    22 & SNR & Signal-to-noise ratio of the horizontal/vertical component of the pen position \\
    \hline
    23 & EMD & Empirical Mode Decomposition\\
    \hline
    \end{tabular}
\end{table*}

According to the Table \ref{tab:Baseline features} the 23 kinematic features used in this study encompass a range of metrics that capture both the spatial and temporal aspects of handwriting movement. These features provide a comprehensive view of the pen's behavior during writing tasks, including its position, displacement, and velocity in both horizontal (x) and vertical (y) directions. Features such as pressure, azimuth, and altitude give additional insights into the pen's interaction with the writing surface, reflecting how PD patients may exhibit altered pen grip, pressure variability, and orientation during handwriting. In addition, the number of changes in velocity, acceleration, and pressure track the irregularities in handwriting motion, which may be indicative of tremors, motor control deficiencies, or inconsistencies in movement.

Temporal aspects of the writing process, such as stroke number, durations, and the ratio of in-air time to on-surface time, further enhance the analysis by providing insights into the pacing and fluidity of handwriting. These features, when combined with measures like energy, entropy, and signal-to-noise ratio, allow for a more nuanced understanding of the motor impairments that can occur in PD. In particular, features related to entropy and SNR help quantify the complexity of movement and the influence of noise, while energy captures the intensity of pen motion, which may decrease in PD patients due to reduced motor vigour. These kinematic features, together with their corresponding statistical measures, serve as a powerful set of tools for distinguishing between normal and abnormal handwriting patterns in PD, providing a clearer picture of the subtle motor dysfunctions that may characterize the disease. These baseline features provide a general understanding of handwriting patterns but may lack sensitivity to more nuanced variations, particularly those that can distinguish between PD patients and HC. To over come the gap we proposed new 65 kinematic features. 
\subsubsection{Statistical Feature Extraction From Each Kinamatic Features}

In this study, we extracted a total of 88 kinematic features, of which 65 are newly proposed and 23 are from existing studies. To capture subtle details of each task and movement pattern, we derived $11 \pm 3$ statistical features from each of the kinematic features. These statistical features can be categorized into two types: Central Tendency and Dispersion Features (as described in \ref{feature_statistic_central}), and Higher-order and Relation Statistical Features (as described in \ref{feature_statistic_high_order}). A brief description of the methods used for extracting these statistical features is provided below.

 \paragraph{Central Tendency and Dispersion Features} \label{feature_statistic_central}
 The extraction of Central Tendency and Dispersion Features plays a crucial role in enhancing the representation of a tabular dataset for the recognition of PD features. Features such as mean, median, variance, and standard deviation provide foundational insights into the central values and variability of the data, enabling a robust understanding of normal and anomalous patterns in the dataset \cite{miah2017motor_miah_SVM_PCA_ANOVA,miah2020motor_pca_Anova_LDA_mad,miah2019motor_LDA_MAD}. Extremes are captured using maximum and minimum values, while the 1st and 99th percentiles, combined with the displacement between them, offer a granular view of data spread and outlier behaviour. This detailed statistical profiling helps identify subtle variations in PD-related data, such as slight tremor intensities or movement inconsistencies, which may otherwise be overlooked in raw data formats. As a result, these features enable the model to capture both average trends and deviations effectively, improving its ability to discern patterns specific to PD. Table \ref{tab:statistical_features} demonstrated the feature's name. 
The extraction of meaningful statistical features from the Kinematic features enables an insightful understanding of the dynamics within a signal. Here are the 13 statistical features calculated for each Kinematic features. The statistical feature extraction process starts by transforming raw Kinematic features into statistically rich features that reveal patterns and trends hidden in the signal. Features like mean, median, and RMS encapsulate the central tendency and magnitude, while variance and standard deviation illuminate variability. Maximum and minimum identify extreme behaviours, and percentiles provide thresholds that uncover outliers. The statistical feature and their mathematical definitions and descriptive explanations in Table \ref{tab:statistical_features}:

\begin{table*}[]
    \centering
    \caption{Features and Their Descriptions}
    \label{tab:statistical_features}
    \begin{tabular}{|l|m{300pt}|}
        \hline
       Feature Name & \textbf{Description} \\
        \hline
        \textbf{Mean} & The mean represents the average value of the signal. It provides a central measure of the data, indicating the typical magnitude of the signal over its entire duration. 
        \[
        \text{Mean} (\mu) = \frac{1}{N} \sum_{i=1}^{N} x_i
        \] \\
        \hline
        Median & The median is the middle value of the signal when sorted in ascending order. It is a robust measure against outliers and provides a central distribution value. 
        \[
        \text{Median} = x_{\lceil N/2 \rceil} \quad \text{(if sorted data: } x_1, x_2, \ldots, x_N \text{)}
        \] \\
        \hline
        \textbf{Variance} & Variance quantifies the spread or dispersion of the signal. A higher variance indicates more variability in the signal values. 
        \[
        \text{Variance} (\sigma^2) = \frac{1}{N} \sum_{i=1}^{N} (x_i - \mu)^2
        \] \\
        \hline
       Standard Deviation & Standard deviation is the square root of variance and provides an intuitive measure of the average deviation from the mean. 
        \[
        \text{SD} (\sigma) = \sqrt{\frac{1}{N} \sum_{i=1}^{N} (x_i - \mu)^2}
        \] \\
        \hline
        Maximum & The maximum is the highest value within the signal. It identifies peaks or extreme values in the data. 
        \[
        \text{Max} = \max(x_1, x_2, \ldots, x_N)
        \] \\
        \hline
        Minimum & The minimum is the lowest value in the signal. It captures the valleys or troughs of the data. 
        \[
        \text{Min} = \min(x_1, x_2, \ldots, x_N)
        \] \\
        \hline
        1st Percentile & The 1st percentile is the value below which 1\% of the data falls. It provides insight into the lower extremes of the distribution. \\
        \hline
        99th Percentile & The 99th percentile is the value below which 99\% of the data falls. It highlights the upper extremes of the distribution. \\
        \hline
        Displacement Between 99th and 1st Percentiles & This feature measures the range between the upper and lower extremes of the data. It represents the spread of the most extreme values. 
        \[
        Displacement = \text{Percentile}_{99} - \text{Percentile}_{1}
        \] \\
        \hline
        Skewness & Skewness quantifies the asymmetry of the data distribution. Positive skew indicates a longer tail to the right, while negative skew indicates a longer tail to the left. 
        \[
        \text{Skewness} = \frac{\frac{1}{N} \sum_{i=1}^N (x_i - \mu)^3}{\sigma^3}
        \] \\
        \hline
        Kurtosis & Kurtosis measures the "tailedness" of the data distribution. Higher kurtosis signifies more extreme outliers. 
        \[
        \text{Kurtosis} = \frac{\frac{1}{N} \sum_{i=1}^N (x_i - \mu)^4}{\sigma^4} - 3
        \] \\
        \hline
        % \hline
        % Root Mean Square (RMS) & RMS is a measure of the magnitude of a varying signal. It gives a sense of energy or intensity. 
        % \[
        % \text{RMS} = \sqrt{\frac{1}{N} \sum_{i=1}^N x_i^2}
        % \] \\
        % \hline
        % Correlation Coefficient & The correlation coefficient measures the strength and direction of the relationship between two signals. 
        % \[
        % \text{CorrCoeff} = \frac{\text{Cov}(x, y)}{\sigma_x \sigma_y}
        % \] \\
        % \hline
    \end{tabular}
\end{table*}

%Two types of statistical feature we extracted here: mean, median, variance, standard deviation, maximum ,minimum, 1st percentile, 99th percential, displacement between 99th and 1st percentile. 2nd included skewness, kurtosis, root mean square, correlation coefficient. 
 \paragraph{Higher-Order and Relationship}\label{feature_statistic_high_order}
The Higher-Order and Relationship Features capture the shape of data distributions and inter-variable relationships, enhancing the analysis. Skewness and kurtosis are key metrics: skewness indicates asymmetry in the data, while kurtosis measures the "peakedness" or presence of extreme values. These features are crucial for identifying abnormalities in PD patients, such as uneven gait or spastic movements, improving model sensitivity to subtle disease markers. As shown in Table \ref{tab:statistical_features}, these features enhance classification accuracy by identifying complex patterns in PD datasets.

Kurtosis and skewness offer deeper insights than traditional statistical measures. Kurtosis reveals the sharpness of data peaks, highlighting extreme values, while skewness indicates the direction of data asymmetry. These higher-order features, applied to handwriting movement data, allow more effective differentiation between HC compared to standard statistical measures. Additionally, we propose Signed Speed Features, which capture movement direction information for speed-related features, particularly those involving the direction of an arbitrary axis. After extracting a comprehensive set of statistical features from each kinematic feature, we calculated a total of 944 features, which were then fed into the subsequent stages of analysis.

\subsection{Feature Normalization}
All data used in training and validation are properly standardized prior to experimentation with machine learning models. More specifically, standardization was performed using StandardScaler and Pipeline, which were provided in the scikit-learn library. The training data mean and standard deviation were calculated, and the training and validation data were standardized using the training data mean and standard deviation.
\begin{equation}
     \text{Normalized Value} = \frac{x - \mu}{\sigma}
\end{equation}
where $\mu$ is mean value, $\sigma$ is standard deviation.

\subsection{Feature Selection}
In general, feature selection suppresses overfitting and increases versatility, as well as reduces computational costs and shortens training time.  In the study, we selected features in two ways: Top $K\%$ features selection and SFFS. In handwriting-based tabular data analysis for dynamic movement statistical feature extraction, the selection of relevant features is critical to the performance of machine learning algorithms. High-quality features significantly enhance the accuracy and reliability of predictions, particularly in tasks related to the classification and detection of PD patients.

To optimize the selection process, we employed two feature selection techniques: RF and SFFS, targeting the most influential features from the handwriting data. Initially, an RF-based algorithm was utilized to rank features based on Gini impurity scores, identifying the top N\% of features as the most important. The RF algorithm's ability to rank features by their importance makes it a valuable tool for reducing dimensionality while preserving the most relevant data for further analysis.

\subsection{Top $K\%$ Feature Selection}

Random Forest (RF) is a supervised learning algorithm combining decision trees and ensemble learning, effective for classification and feature selection. In our study, we used an RF model to identify top-ranking features through the following steps:

\begin{itemize} 
\item $N$ subsamples were created from the original data using bootstrap sampling (random sampling with replacement), increasing model diversity. \item $N$ decision trees were generated, where at each node, $M$ features were randomly selected for splitting, reducing overfitting. \item The RF model was trained with combined features. \item Gini impurity was calculated for each feature, ranking them based on importance in classification. \item The top 50\%, 25\%,  10\%, 5\%  and 1\% of features, based on Gini impurity scores, were selected for further analysis. \end{itemize}

This approach effectively reduced the dimensionality while retaining the most informative features for handwriting-based analysis, aiding in PD identification.

\subsection{Sequential Floating Forward  Selection}

In the study, we employed SFFS to select the potential features, which are shown in Figure \ref{fig:sffs}. SFFS is mainly a type of greedy algorithm to select the effective feature aiming to reduce dimensionality, which is widely used to select a subset of $k$-dimensional features from a $d$-dimensional feature space (where $k < d$) and plays a crucial role in reducing generalization errors \cite{SFFS}. In this study, we employed SFFS to select relevant features for PD patients across the various tasks for the following reasons: (i) This method employs the strengths of both forward and backward elimination to identify the most informative features, minimizing redundancy in the process; (ii) it initializes with an empty set, adding features incrementally to maximize model performance and interpretability, and stops once no further enhancements are achieved. The final selected feature subset is applied in ML-based gender classification (PD/HC). This approach also significantly enhances model interpretability and performance and reduces computational costs. In this experiment, we chose the Sequential Forward Selection(SFS)\cite{SFFS} of the Wrapper Method and the pseudocode of the SFFS is given below: 

\begin{algorithm}
    \caption{SFFS Algorithm \cite{SFFS}}
    \begin{algorithmic}[1]
        \STATE \textbf{Input:} $Y = \{y_1, y_2, \ldots, y_d\}$: set of all features
        \STATE $X_k = \{x_j \mid j = 1, 2, \ldots, k; x_j \in Y\}$, where $k \in \{0, 1, 2, \ldots, d\}$ and $X_k$ is a subset of $Y$
        \STATE \textbf{Output:} Selected feature subset $X_k$
        \STATE \textbf{Initialization:} $X_0 = \emptyset$, $k = 0$
        \REPEAT
            \STATE $x^{+} = \arg\max J(X_k + x)$, where $x \in Y \setminus X_k$, $J$ is an evaluation index, and $x^{+}$ is the feature with the highest evaluation when selected.
            \STATE $X_k = X_k + x^{+}$
            \STATE $k = k + 1$
        \UNTIL $k$ reaches the specified number of features
        \STATE \textbf{Step 1} to \textbf{Step 3} are repeatedly iterated. When $k$ reaches the specified number, $x^{+}$ is the set of the most appropriate features obtained. SFFS is performed up to \textbf{Step 3} of SFS, and a process for searching for features to be deleted is added. Initially, \textbf{Step 1} to \textbf{Step 4} are performed starting from $X_0 = \emptyset$, $k = 0$, as in the SFS.
        \STATE $x^{-} = \arg\max J(X_k - x)$, where $x \in X_k$ and $x^{-}$ is the feature with the best performance when the feature is deleted.
        \IF {$J(X_k - x) > J(X_k)$}
            \STATE $X_k = X_k - x^{-}$
            \STATE $k = k - 1$
            \STATE Go to \textbf{Step 1}
        \ENDIF
    \end{algorithmic}
\end{algorithm}

\Figure[htbp](topskip=0pt, botskip=1pt, midskip=0pt)[width=.40\textwidth]{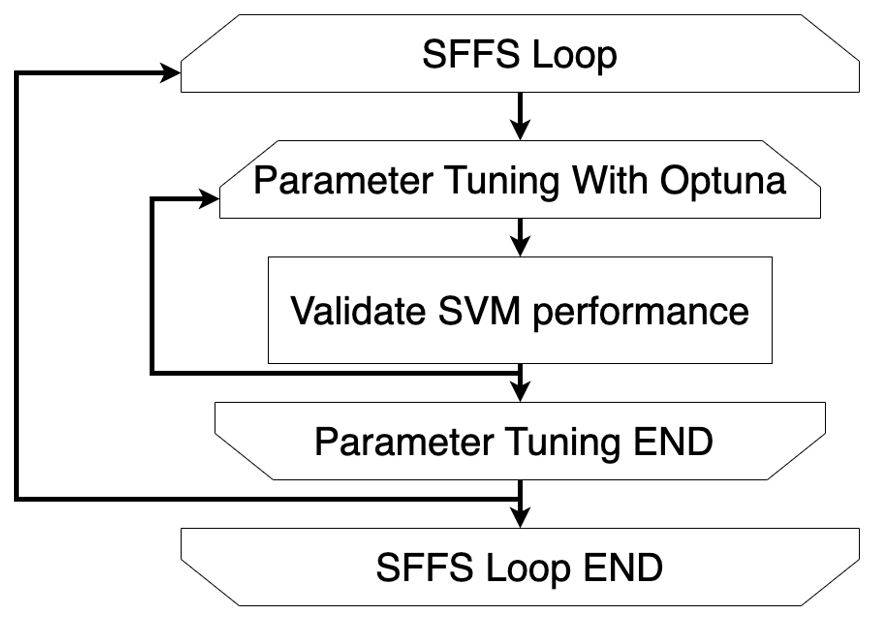}{SFFS architecture.\label{fig:sffs}}

\subsection{Machine Learning Based Classifier Analysis}

To classify PD and HC, we developed SVM \cite{SVM}  classifiers trained on specific statistical features. We calculated 65 novel and 23 existing statistical features for each of the seven tasks, which include measurements pertinent to Parkinson’s diagnosis. Each task-specific SVM classifier is optimized for performance, with parameters fine-tuned via Bayesian optimization using the Optuna library \cite{Optuna}. 

\subsubsection{SVM Classifier and Kernel Optimization}
The SVM classifier \cite{miah2022movie_miah_SVM,kabir2023investigating_miah_SVM_LDA_MLP} seeks to find an optimal hyperplane to separate PD and HC classes, given by:
\begin{equation}
f(\mathbf{x}) = \text{sign} \left( \sum_{i=1}^{n} \alpha_i y_i K(\mathbf{x}_i, \mathbf{x}) + b \right)
\label{eq:svm_classifier}
\end{equation}
where \( \mathbf{x}_i \) are support vectors, \( y_i \) are their class labels, \( \alpha_i \) are Lagrange multipliers, \( K(\mathbf{x}_i, \mathbf{x}) \) is the kernel function, and \( b \) is the bias term. We explored several kernels:
\begin{itemize}
    \item \textbf{Linear kernel}:
    \begin{equation}
    K(\mathbf{x}_i, \mathbf{x}) = \mathbf{x}_i \cdot \mathbf{x}
    \label{eq:linear_kernel}
    \end{equation}

    \item \textbf{RBF kernel}:
    \begin{equation}
    K(\mathbf{x}_i, \mathbf{x}) = \exp \left( -\gamma \|\mathbf{x}_i - \mathbf{x}\|^2 \right)
    \label{eq:rbf_kernel}
    \end{equation}

    \item \textbf{Sigmoid kernel}:
    \begin{equation}
    K(\mathbf{x}_i, \mathbf{x}) = \tanh(\alpha \mathbf{x}_i \cdot \mathbf{x} + c)
    \label{eq:sigmoid_kernel}
    \end{equation}
\end{itemize}
The hyperparameters \( C \) and \( \gamma \) were optimized within the range \( 0.01 \) to \( 100 \), and the Leave-One-Out cross-validation method was used to evaluate generalization performance for each task classifier.

\subsubsection{Individual Task Classification and Ensemble Approach}
We trained an optimized SVM for each of the 7 tasks, each leveraging a different set of statistical features. After calculating the accuracy of each task-specific classifier, we proceeded to ensemble combinations of these classifiers to enhance overall classification performance.
\begin{itemize}
    \item \textbf{Top-3 Task Ensemble}: We selected the top three task-specific SVMs with the highest individual accuracy. The ensemble output \( F_{3}(\mathbf{x}) \) is given by:
    \begin{equation}
    F_{3}(\mathbf{x}) = \text{sign} \left( \sum_{i \in T_3} w_i f_i(\mathbf{x}) \right)
    \label{eq:top3_ensemble}
    \end{equation}
    where \( T_3 \) represents the indices of the top-3 performing tasks, \( f_i(\mathbf{x}) \) is the classifier output for the \( i \)-th task, and \( w_i \) are weights typically based on individual accuracy scores.

    \item \textbf{Top-5 Task Ensemble}: We extended the ensemble to include the top five task-specific SVMs. The ensemble output \( F_{5}(\mathbf{x}) \) is given by:
    \begin{equation}
    F_{5}(\mathbf{x}) = \text{sign} \left( \sum_{i \in T_5} w_i f_i(\mathbf{x}) \right)
    \label{eq:top5_ensemble}
    \end{equation}
    where \( T_5 \) contains the indices of the top-5 performing tasks.
    \item \textbf{All-Task Ensemble}: Finally, we created an ensemble using all 7 task-specific SVMs, yielding the output \( F_{7}(\mathbf{x}) \):
    \begin{equation}
    F_{7}(\mathbf{x}) = \text{sign} \left( \sum_{i=1}^{7} w_i f_i(\mathbf{x}) \right)
    \label{eq:all_task_ensemble}
    \end{equation}
    Here, each \( w_i \) represents the weight for the classifier of the \( i \)-th task, which can be set based on validation accuracy or treated uniformly.
\end{itemize}
Each ensemble approach was validated using Leave-One-Out cross-validation to ensure robust performance. The results showed that the ensemble methods improved classification accuracy, especially when aggregating outputs from individually optimized tasks. Fig. \ref{fig:svm_ensemble} shows the task-wise ensemble ideas.  Equations \ref{eq:svm_classifier} through \ref{eq:all_task_ensemble} describe the mathematical formulations of the SVM classifiers and ensemble methods used in our analysis. 

%We used Support Vector Machine(SVM)\cite{SVM} to classify. However, SVM has multiple parameters, and it is known that tuning can improve performance. Therefore, by performing parameter tuning while evaluating performance, optimized parameters for each feature value can be used, and features can be more fairly selected. The evaluation of features is based on the classification accuracy in cross-validation by the Leave-One-Out method. Parameter Tuning performed a search by Bayesian optimization using the Optuna library \cite{Optuna}. The kernel was searched for ["linear", "rbf", "sigmoid"], and $C$ and $gamma$ were searched in the range 0.01 \textasciitilde 100. Classify PD and non-PD using the features and tuned parameters selected in the section above. Most of the time, it is known that the more features you choose, the higher the accuracy. However, this experiment did not improve accuracy compared to using only one feature to the combination of two or more. Details are given in the results column. Therefore, we have multiple SVMs trained using only one feature and ensemble their outputs to perform PD/HC classification. Even in this case, cross-validation using the Leave-One-Out method evaluates the generalization performance of this method. The advantage of this method is that we can choose feature vectors optimized for classification for each task and ensemble them.

\Figure[htbp](topskip=0pt, botskip=1pt, midskip=0pt)[width=.70\textwidth]{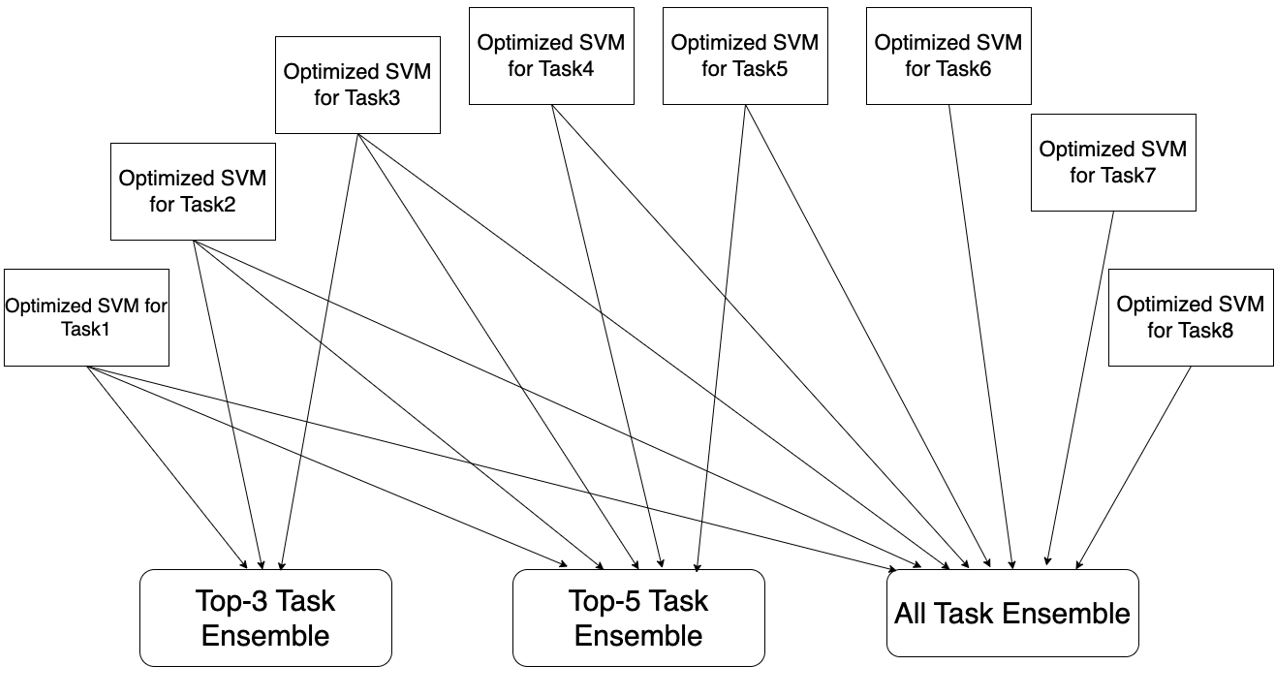}{Ensemble Tasks Classification.\label{fig:svm_ensemble}}

\section{Experimental Results}

The experimental evaluation was conducted using the PaHaW dataset, which consisted of 37 PD patients and 38 HC. Each participant completed a series of eight handwriting tasks designed to capture a range of motor skills. These tasks included: 1) drawing the Archimedes spiral; 2) writing the cursive letter “l”; 3) writing the bigram “le”; 4) writing the trigram “les”; 5) writing the cursive word “lektorka” (meaning “female teacher” in Czech); 6) writing the word “porovnat” (which means “to compare”); 7) writing “nepopadnout” (meaning “to not catch”); and 8) composing the cursive sentence “Tramvaj dnes už nepo-jede” (translating to “The tram won’t go today”). To ensure the robustness of the classification results, all extracted features were normalized to achieve a zero mean and unit variance. For the purpose of evaluating model performance, we implemented Leave-One-Out Cross-Validation (LOOCV). 

\subsection{Leave-One-Out Cross-Validation}

LOOCV is a specific type of cross-validation technique used to assess the performance of a predictive model. In this approach, the dataset is divided into multiple subsets, where each subset consists of a single observation (data point). The model is trained on all data points except for the one left out and then tested on that single observation. This process is repeated for each observation in the dataset, resulting in as many iterations as there are observations \cite{miah2023dynamic_graph_general,miah2024sign_largescale,shin2023korean_ksl1}.
The primary advantage of LOOCV lies in its ability to utilize nearly the entire dataset for training, which can lead to a more accurate estimation of the model's performance. However, it is computationally intensive, especially with larger datasets, since it requires training the model multiple times. Despite this, LOOCV can provide valuable insights into how well the model generalizes to unseen data, making it a useful technique for evaluating the performance of classifiers in studies like ours.

\subsection{Experimental Setup and Performance Metrics}

This section presents the details of the experimental environment and Python/libraries.
This experiment was run on a PC (with Ubuntu 22.04, Intel Core i9-13900K, 64GM RAM, Python:3.10.4, scikit-learn:1.3.0, Optuna:2.10.1, Numpy:1.23.1, Pandas:1.4.3).
The performance of the trained classification models was evaluated by classification accuracy (ACC), recall, precision, and F1-Score (F1S) which are defined as follows:

\begin{equation}
\text{ACC} = \frac{TP + TN}{TP + TN + FP + FN} \times 100[\%]
\end{equation}

\begin{equation}
\text{Recall} = \frac{TP}{TP + FN} \times 100[\%]\\
\end{equation}

\begin{equation}
\text{Precision} = \frac{TP}{TP + FP} \times 100[\%]\\
\end{equation}

\begin{equation}
\text{F1S} = 2 \times \frac{\text{Precision} \times \text{Recall}}{\text{Precision} + \text{Recall}}
\end{equation}

where TP is the number of true positives, TN is the number of true negatives, FP is the number of false positives, and FN is the number of false negatives.

\subsection{Ablation Study}

In this study, we introduced novel dynamic movement statistical features and combined them with existing baseline features to form a comprehensive feature vector. To ensure the selection of the most relevant features, we applied the SFFS-based algorithm across both task-specific and ensemble tasks, leveraging state-of-the-art ML-based algorithms. For the ablation study, we further refined RF model for feature selection to rank the top k\% features, and evaluated model performance using 10-fold CV. We conducted experiments by progressively selecting the top 50\%, 25\%, 10\%, 5\%, and 1\% of features and comparing the accuracy across various algorithms, using both k-fold and LOOCV splitting techniques. This approach highlights the robustness of our model by revealing how feature reduction impacts performance, ensuring that the selected features offer optimal accuracy with minimal computational cost. The novelty lies in combining new dynamic features with advanced feature selection methods, demonstrating the necessity of efficient feature extraction and selection in improving classification performance.

\subsubsection{Ablation Study using All Features and RF Feature Selection with with k-Fold Cross Validation}

Table \ref{tab:ablation_kcv} shows the ablation study, whereas we applied k-fold CV to evaluate the performance of various state-of-the-art algorithms, using the top k\% selected features. The study showed how feature selection impacts model performances across different algorithms. For the SVM model, the performance improved as feature selection becomes more refined, reaching 64.29\% with 1\% of the top features. Extra Trees showed fluctuating results, peaking at 66.79\% with 5\% of features, demonstrating its sensitivity to the number of selected features. Random forest (RF) achieved the highest accuracy of 68.39\% with 1\% of features, highlighting its robustness in feature reduction. Gradient Boosting performed consistent improvements, with the best performance at 73.75\% using 25\% of features. AdaBoost remained stable, reaching 73.57\% in both full features and 1\% of the top selected features, indicating its resilience. Lastly, KNN-based model showed modest improvements, with the highest accuracy of 55.89\% using 5\% of selected features.

\begin{table*}[ht!]
\centering
\caption{Performance (in \%) of the Ablation study with K-fold CV using all features and RF feature selection } \label{tab:ablation_kcv}

\begin{tabular}{|c|c|p{2cm}|c|c|c|c|}
\hline
\textbf{Algorithms} & \textbf{All Features} & \textbf{RF\_Selection\_50\% of Features} & \textbf{25\% of Features} & \textbf{10\% of Features} & \textbf{5\% of Features} & \textbf{1\% of Features} \\ 

\hline
SVM   & 61.43        & 61.25              & 61.79    & 63.04    & 63.04    & 64.29    \\ 

\hline
ET        & 62.68        & 64.11        & 64.29    & 61.07    & 66.79    & 58.39    \\ 
\hline

RF    & 64.11        &     -     & 65.71    & 62.86    & 60.18    & 68.39    \\ 
\hline
GB   & 69.64        & 72.86        & 73.75    & 70.00    & 68.39    & 67.14    \\
\hline
AB        & 73.57        & 70.89        & 72.32    & 73.57    & 72.32    & 73.57    \\ 
\hline
KNN     & 53.04        & 53.04        & 51.79    & 55.71    & 55.89    & 54.11    \\ 
\hline
\end{tabular}
\end{table*}

%hirooka shared file; https://docs.google.com/spreadsheets/d/1JZzn47sicsBYwBY7ycBnXFhS6yzkTJ9yOffm2I0DYdE/edit?gid=0#gid=0 

\subsubsection{Ablation Study using All Features and RF Feature Selection with LOOCV}

In this ablation study, LOOCV was employed to evaluate the performance of various algorithms across different levels of feature selection, as illustrated in Table \ref{tab:ablatio_loocv}. The results highlight how selecting a percentage of the top features influences classification accuracy for different algorithms.
SVM achieved the highest accuracy of 66.67\% when selecting the top 10\% of features, indicating an optimal balance between feature reduction and model performance. The Extra Trees algorithm displayed consistent performance across all feature levels, peaking at 66.67\% accuracy when all features were included, showing robustness to feature selection.
Gradient Boosting attained a classification accuracy of 88.89\% when using the top 50\% of features and 84.72\% with all features. However, its accuracy dropped with fewer features, reaching 66.67\% when only 1\% of features were selected. AdaBoost also performed well, achieving 77.78\% classification accuracy when selecting 50\% of the features. In contrast, the KNN model exhibited modest performance compared to other algorithms, with its highest classification accuracy of 55.56\% for 50\% feature selection. Its performance remained stable but consistently lower than that of the other models.
Overall, Gradient Boosting and AdaBoost showed improved performance with an increase in the number of selected features, while SVM and Extra Trees exhibited greater resilience to feature reduction. KNN, although less sensitive to feature selection, consistently achieved lower accuracy. These findings underscore the critical role of feature selection in enhancing model performance.

\begin{table*}[ht!] 
\centering
\caption{Performance (in \%) of the Ablation study with LOOCV using all features and RF feature selection}
\label{tab:ablatio_loocv}
\begin{tabular}{|c|c|p{2cm}|p{2cm}|p{2cm}|p{2cm}|p{2cm}|}
\hline
\textbf{Algorithms} & \textbf{All Features} & \textbf{Selected\_Top\_50\% Features} & \textbf{Selected Top 25\% Features} & \textbf{Selected Top 10\% Features} & \textbf{Selected Top 5\% Features} & \textbf{Selected Top 1\% Features} \\ 
\hline

SVM              & 63.89        & 62.50              & 65.28    & 66.67    & 65.28    & 63.89    \\ \hline
ET       & 66.67        & 62.50        & 63.89    & 61.11    & 62.50    & 62.50    \\ \hline
GB        & 84.72             & 88.89             & 86.11             & 70.83             & 73.61             & 66.67             \\ \hline
AB              & 76.39             & 77.78             & 73.61             & -                 & 77.78             & -                 \\ \hline
KNN                    & 54.17             & 55.56             & 54.17             & 54.17             & 52.78             & 54.17             \\ \hline

\end{tabular} \\
SVM: Support Vector Machine; ET: Extra Tree; GB: Gradient Boost; AB: Ada Boost. 
\end{table*}

\subsection{Performance Result with With Selected Features}

We trained SFFS-SVM model with LOOCV protocol and optimized their parameters using Optuna while changing the feature-subsets used. We identified the optimized feature subsets using classification accuracy and their correspondence results are presented in Table \ref{tab:Selected_Feat_Taskwise}. This table highlights the effectiveness of various features for predicting PD patients in various tasks, with most tasks demonstrating high precision, recall, and F1 scores. We observed that the best classification accuracy was obtained for one feature in most of the tasks. The classification accuracy did not improve by considering two or more features. Each row outlined a specific task, detailing the feature name used and the performance metrics achieved. Notably, Task 1 achieved the highest accuracy of 98.61\%. Whereas, Task 6 recorded the lowest accuracy of 93.33\%. 

\begin{table*}

\centering
    \caption{Classification accuracy (in \%) of the selected features for each task}
    \label{tab:Selected_Feat_Taskwise}
    \begin{tabular}{|c|p{2cm}|p{6cm}|c|c|c|c|}
    \hline
    Task Number &Task Name& Selected Features & ACC & Precision & Recall & F1 \\
    \hline
    Task 1 & spiral&\text{Teager\_Kaiser\_Energy(velocity,1)} & 98.61 & 98.65 & 98.61 \% & 98.61 \\
    
    Task 2& "lll" & \text{first\_x\_1st\_percentile} & 98.66 & 98.72 & 98.65 & 98.67 \\

    Task 3 & "le le le"& \text{last\_y\_signed\_velocity\_min} & 98.66 & 98.68  & 98.68  & 98.67 \\
    
    Task 4 & "les les les"& \text{last\_x\_displacement\_kurtosis, SNRce(displacement)} & 96.0 & 96.05 & 95.98  & 96.00 \\
    
    Task 5 & "lektorka"& \text{first\_displacement\_max, first\_x\_acceleration\_max} & 98.66 & 98.72 & 98.65 & 98.67 \\
    
    Task 6 & "porovnat"& \text{first\_x\_signed\_displacement\_kurtosis} & 93.33 & 93.65 & 93.28  & 93.31 \\
    
    Task 7 & "nepopadnout"& \text{velocity\_kurtosis} & 94.66 & 94.82 & 94.63 & 94.66 \\
    
    Task 8 & "tramvaj dnes uz nepo-jede"& \text{first\_displacement\_skewness} & 93.33 & 93.65  & 93.28  & 93.31 \\
    \hline
    \end{tabular}

\end{table*}

\subsection{State of the Art Comparison}
We performed a comparison study of our proposed method against previous models published in two ways: task-wise and combined task with sample-wise which are briefly explained the following sub-subsection \ref {Task-Wise} and 
sub-subsection \ref {Combined Task-Wise}.

\subsubsection{Task-wise State of the Art Comparison}
\label{Task-Wise}

Table \ref{tab:Comparison_of_Classification_Accuracy_each_task} highlights the efficacy of our proposed method in comparison to existing approaches, particularly in maintaining high classification accuracy across multiple tasks. As indicated in Table \ref{tab:Comparison_of_Classification_Accuracy_each_task}, Peter et al. \cite{related-1} proposed a PD detection model and achieved an average classification accuracy of 69.25\%, with the highest accuracy of 79.4\% for Task 8, whereas the lowest classification accuracy of 65.4\% for Task 1 \cite{related-1}. In contrast, Donato Impedovo \cite{related-4} demonstrated superior performance, achieving an average accuracy of 95.57\% and the best detection rate of  97.33\% for Task 1 whereas Task 4 provided the lowest accuracy of 93.13\% \cite{related-4}. Our proposed method outperformed compared to existing models \cite{related-1, related-4}. More specially, our proposed method obtained an average classification accuracy of 96.99\%, with the highest accuracy of 98.66\% in both Task 2 and Task 3, while the lowest classification of 93.33\% was obtained for both Tasks 6 and Task 8. 

\begin{table*}[ht!]
\centering
    \caption{Task-wise accuracy (in \%) comparison of proposed system against state of the art model}
    \label{tab:Comparison_of_Classification_Accuracy_each_task}
    \begin{tabular}{|c||c|c|c|c|c|c|c|c|c|}
    \hline
    Method & Task1 & Task2 & Task3 & Task4 & Task5 & Task6 & Task7 & Task8 & Average \\
    \hline
    Peter et.al,\cite{related-1} & 65.4 & 70.0 & 72.3 & 65.4 & 66.7 & 67.7 & 67.1 & 79.4 & 69.25 \\
    Impedovo\cite{related-4} & 97.33 & 97.43 & 95.12 & 93.13 & 96.79 & 95.96 & 96.76\% & 92.05 & 95.57 \\
    \hline
    Proposed Method & 98.61 & 98.66 & 98.66 & 96.00 & 98.66 & 93.33 & 94.66 & 93.33\% & \textbf{96.99} \\
    \hline
    \end{tabular}
\end{table*}

\subsubsection{Combined Task-wise State of the Art Comparison}
\label{Combined Task-Wise}

Table \ref{tab:Comparison_Ensemble_Classification_Accuracy_with_existing_study} compares the state-of-the-art performance of various models, including the proposed model, across combined tasks, focusing on accuracy and methodologies. Each column represents critical aspects of the approaches, such as the method used, the number of combined tasks, the names of those tasks, the feature types extracted (along with their total number), the feature selection algorithm utilized (with the count of selected features), the classification algorithm applied, and the accuracy achieved. This result outperforms the current state-of-the-art model. Peter et al. explored different task combinations and achieved classification accuracies ranged from 79.4\% to 88.1\%, obtained by SVM and various feature selection methods \cite{related-1,related-2,related-3}. Mucha et al. also utilized advanced featuree and obtained a classification accuracy of 97.14\% using XGBoost \cite{related-5}. Impedovo et al. showed performance using both baseline and newly proposed features and obtained the classification accuracies between 88.33\% and 97.14\% \cite{related-4}. In contrast, the proposed method demonstrates superior performance, achieving an impressive accuracy of 99.98\% across various task combinations. This is accomplished by employing a combination of baseline and custom dynamic features. The important features were selected by the SFFS-based algorithm. The use of SVM for classification further enhances the robustness of the proposed methodology, highlighting its strength in effectively distinguishing complex patterns within handwriting data for PD detection.

\begin{table*}[ht!]
\small
\centering
    \caption{State of the Art Accuracy (in \%) Comparison of Combined Tasks of The Proposed Model by Combining Different Tasks Samples}
    \label{tab:Comparison_Ensemble_Classification_Accuracy_with_existing_study}
    \begin{tabular}{|c|c|p{2cm}|p{3cm}|p{2cm}|c|c|}
    \hline
    Authors & Combined Tasks & Combined Task Name & Extracted Features (Total No) & FSA (Selected Feature No) & CA & ACC \\
    \hline
    Peter et.al,\cite{related-1} & 8 & Task 1, 2, 3, 4, 5, 6, 7, 8 & Trajectory, Velocity, Dynamic, Acceleration, Jerk, Stroke (13) & None & SVM & 79.4 \\ 
    \hline
    
    Peter et.al,\cite{related-2} & 8 & Task 1, 2, 3, 4, 5, 6, 7, 8  & Trajectory, Velocity, Dynamic, Acceleration, Jerk, Stroke , (600) & Whitney U-test filter and relief algorithm (16) & SVM & 80.09 \\
    \hline
    Peter et.al,\cite{related-3} & 7 & Task  2, 3, 4, 5, 6, 7, 8  &Entropy, Signal energy,EMD   & Mann-Whitney U-test filter and relief algorithm  & SVM  & 88.1 \\ 
    
    \hline
    Peter et.al,\cite{related-7} & 7 & Task 1, 2, 3, 4, 5, 6, 7 & Entropy, EMD & None & SVM & AUC-89.09 \\ 
    \hline
    
    Mucha et.al,\cite{related-5} & 8 & Task 1, 2, 3, 4, 5, 6, 7, 8 & Velocity, Acceleration, Frequency Domain,Fractional-Order Derivatives & - &XGBoost & 97.14 \\ \hline
    
    Impedovo,\cite{related-4} & 8&Task 1, 2, 3, 4, 5, 6, 7, 8  & Baseline, 24 & - & SVM & 88.33 \\
    \hline
    
    Impedovo,\cite{related-4} & 8&Task 1, 2, 3, 4, 5, 6, 7, 8  & Baseline+New, 24+3 & - & SVM & 93.79 \\
    \hline

    Impedovo,\cite{related-4} & 3&Task 1, 2, 5  & Baseline+New, 24+3 & - & SVM & 97.14 \\
    \hline
    
    Proposed (Top-3) & 3 & Task 1, 2, 3 & Baseline+ Proposed Dynamic (38) & SFFS (Custom) & SVM & 99.98 \\
    Proposed (Top-5) & 5 & Task 1, 2, 3, 4, 5 & Baseline+ Proposed Dynamic (38) & SFFS (Custom)& SVM & 99.98 \\ 
    Proposed (All) & 8 & All Tasks & Baseline+ Proposed Dynamic (38) & SFFS (Custom) & SVM & 99.98 \\ 
    \hline
    
    \end{tabular} \\
    \footnotesize CA: Classification Algorithm; FSA: Feature Selection Algorithm 
\end{table*}

\section{Discussion}

This study presented an optimized methodology for detecting PD that effectively captures movement dynamics in handwriting tasks. We extracted features from each kinematic feature in two ways: kinematic and statistical based features. In kinematic-based features, we extracted some features, including angle trajectory, signed x/y displacements, velocity, and first/last displacements. Unlike conventional methods that analyze the entire handwriting task, our methodology focuses on dynamic changes within the first and last 10\% of the task. Moreover, we also extracted some statistical features such as kurtosis, skewness, and angular measures. This enhanced the analysis by capturing subtle variations in speed and directional movement. The findings showed that our proposed system achieved a classification accuracy of 96.99\% for tasks-wise evaluation as shown in Table \ref{tab:Comparison_of_Classification_Accuracy_each_task}  and 99.98\% classification accuracy for ensemble tasks as shown in Table \ref{tab:Comparison_Ensemble_Classification_Accuracy_with_existing_study}. This performance accuracy is around 2\% improvement over existing state-of-the-art models.

First, new features were proposed, including the dynamic movement of the hand with the existing baseline features that enhance the feature dimension and create unique patterns. Second, introducing kurtosis and skewness indices proved valuable. These features reflect differences in handwriting speed variability, a known characteristic of PD patients. Whilw healthy subjects maintain relatively stable writing speeds, patients with PD exhibit greater variability, resulting in measurable deviations in distribution and peak sharpness. Lastly, we employed ensemble learning with SVMs trained on single features to mitigate feature interference, a challenge observed when training with multiple feature vectors simultaneously. In SVM model, the optimization of kernels played a crucial role of enhancing the classification accuracy. In our study, we observed that SVM with sigmoid kernel performed better performances compared to other kernels. Whereas, previous studies commonly used RBF or Linear kernels. By using individual SVMs for each feature and combining their outputs through ensemble learning, we ensured that features did not interfere with another. This approach leverages the strengths of each SVM while minimizing the potential for conflicts among features, enhancing the overall performance and robustness of the model. Finally, our methodology outperforms existing models and provides a robust, reliable framework for PD detection. The combination of innovative feature extraction, advanced SVM optimization, and ensemble learning sets a new benchmark in handwriting-based diagnostic tools, offering promising applications for clinical use globally.

\subsection{Implications for Medical Practice}

Our proposed system offers significant potential that will helpful to medical doctors in accurately detecting patients with PD both in Japan and globally. In Japan, where the aging population is increasing the prevalence of neurodegenerative diseases like PD, this system provides a non-invasive, cost-effective, and objective diagnostic tool. By analyzing handwriting data, doctors can detect early signs of PD that might be missed in standard clinical examinations. The highest classification accuracy of this system reduces the risk of misdiagnosis and enables early intervention, improving patient outcomes and reducing the societal burden of the disease. Moreover, the system adaptability to different languages and cultural contexts makes it suitable for deployment in other countries, enhancing global access to PD detection tools. In regions with limited access to advanced neuroimaging facilities or specialized neurological care, this system can bridge diagnostic gaps by providing an affordable and reliable alternative. By equipping medical professionals with actionable insights derived from handwriting analysis, this approach has the potential to revolutionize PD detection and contribute to more personalized and effective patient care worldwide.

\section{Conclusion}

This study proposed an optimized methodology for PD detection that integrated newly developed dynamic kinematic based features with advanced ML-based techniques to capture the movement dynamics during handwriting tasks. By extracting 65 novel kinematic features from the first and last 10\% of the handwriting task, we focused on the critical phases that exhibit significant variations in acceleration, deceleration, and directional changes. This approach not only reduces complexity, but also allows us to capture subtle movements that traditional methods may overlook. Furthermore, we reused 23 existing kinematic features, resulting in a comprehensive set of features designed to improve the accuracy of PD detection. To further refine our approach, we enhance kinematic features using statistical formulas to compute hierarchical-based features, allowing us to better capture subtle movement variations that differentiate patients with PD from HC. We then optimized the feature set by applying the SFFS based method to reduce the dimensionality and computational complexity. Our proposed ML-based ensemble method obtained outstanding results, with a classification of 96.99\% for task-wise evaluations and 99.98\% for task ensemble. This represents a 2\% improvement in classification accuracy compared to the previous models. These outstanding results highlight the potential of our proposed approach to redefine benchmarks for PD detection, offering significant improvements in both classification accuracy and robustness.

In future work, we aim to expand our model by incorporating additional datasets, including both PD patients and healthy controls, to evaluate its generalizability across different populations. Furthermore, we plan to explore real-time deployment of our proposed system, which could provide valuable insights into its practical application for PD diagnosis in clinical settings.

\bibliographystyle{unsrt}
\bibliography{refs}

\begin{IEEEbiography}[{\includegraphics[width=1in,height=1.25in, clip,keepaspectratio]{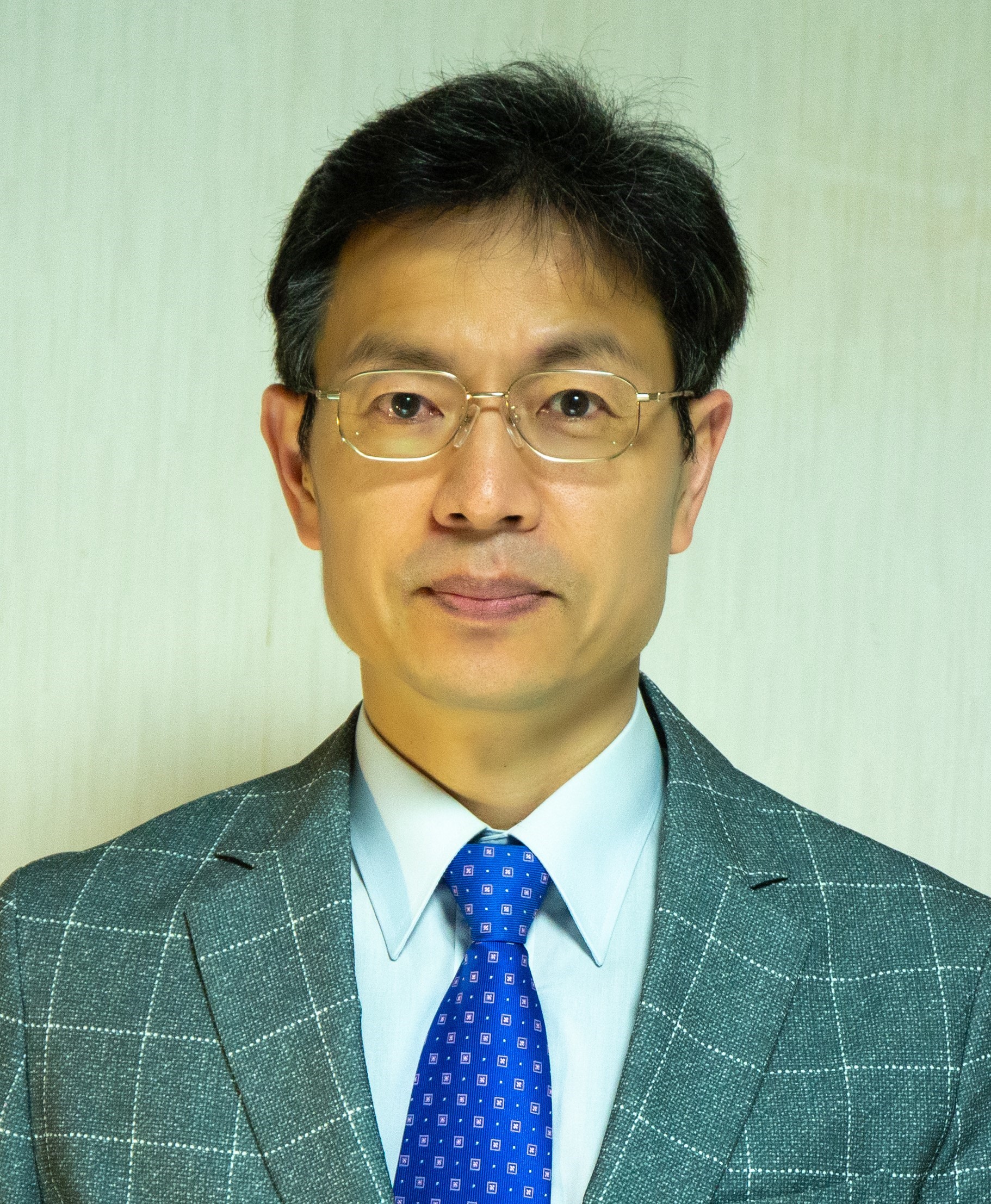}}]
{Jungpil Shin} (Senior Member, IEEE) received the B.Sc. degree in computer science and statistics and the M.Sc. degree in computer science from Pusan National University, South Korea, in 1990 and 1994, respectively, and the Ph.D. degree in computer science and communication engineering from Kyushu University, Japan, in 1999, under a scholarship from the Japanese Government (MEXT). He was an Associate Professor, a Senior Associate Professor, and a Full Professor with the School of Computer Science and Engineering, The University of Aizu, Japan, in 1999, 2004, and 2019, respectively. He has co-authored more than 420 published papers for widely cited journals and conferences. His research interests include pattern recognition, image processing, computer vision, machine learning, human–computer interaction, non-touch interfaces, human gesture recognition, automatic control, Parkinson’s disease diagnosis, ADHD diagnosis, user authentication, machine intelligence, bioinformatics, and handwriting analysis, recognition, and synthesis. He is a member of ACM, IEICE, IPSJ, KISS, and KIPS. He serves as an Editorial Board Member for Scientific Reports. He was included among the top 2\% of scientists worldwide edition of Stanford University/Elsevier, in 2024. He served as the general chair, the program chair, and a committee member for numerous international conferences. He serves as an Editor for IEEE journals, Springer, Sage, Taylor \& Francis, Sensors (MDPI), Electronics (MDPI), and Tech Science. He serves as a reviewer for several major IEEE and SCI journals.
\end{IEEEbiography}

\begin{IEEEbiography}    [{\includegraphics[width=1in,height=1.25in,clip,keepaspectratio]{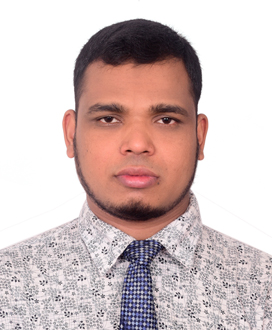}}]
{Abu Saleh Musa Miah} received the B.Sc.Engg. and M.Sc.Engg. degrees in computer science and engineering from the Department of Computer Science and Engineering, University of Rajshahi, Rajshahi-6205, Bangladesh, in 2014 and 2015, respectively, achieving the first merit position. He received his Ph.D. in computer science and engineering from the University of Aizu, Japan, in 2024, under a scholarship from the Japanese government (MEXT). He assumed the positions of Lecturer and Assistant Professor at the Department of Computer Science and Engineering, Bangladesh Army University of Science and Technology (BAUST), Saidpur, Bangladesh, in 2018 and 2021, respectively. Currently, he is working as a visiting researcher (postdoc) at the University of Aizu since April 1, 2024. His research interests include AI, ML, DL, Human Activity Recognition (HCR), Hand Gesture Recognition (HGR), Movement Disorder Detection, Parkinson's Disease (PD), HCI, BCI, and Neurological Disorder Detection. He has authored and co-authored more than 50 publications in widely cited journals and conferences.
\end{IEEEbiography}

\begin{IEEEbiography}
[{\includegraphics[width=1in,height=1.25 in, clip,keepaspectratio]{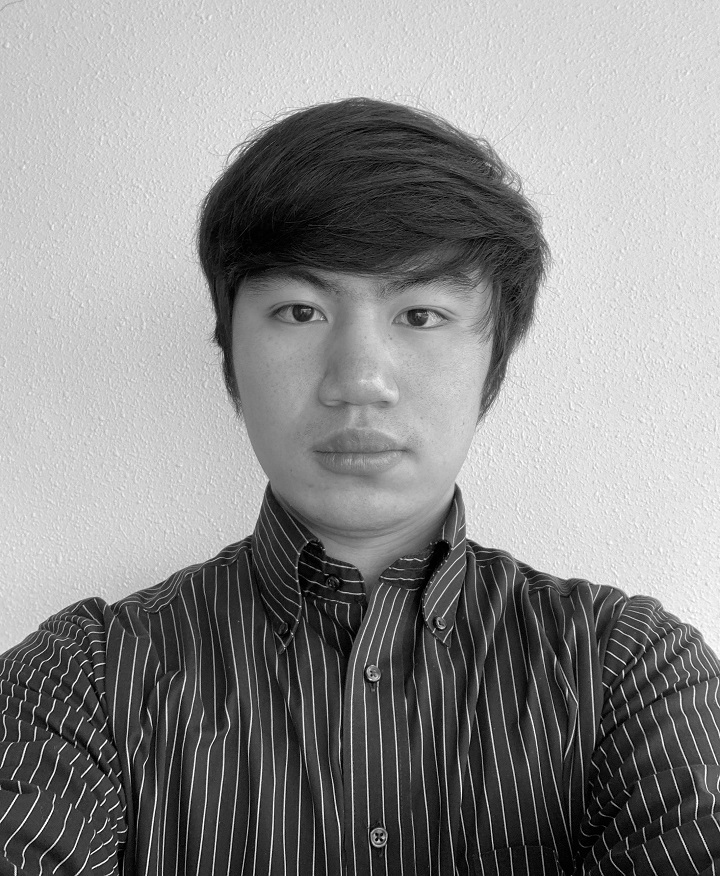}}]{KOKI HIROOKA} was born in Aizumisato-machi,Fukushima, Japan. He received a bachelor's and master's degree in computer science and engineering from The University of Aizu (UoA), Japan, in March 2022 and 2024, respectively. He is currently pursuing a Doctoral degree. He joined the Pattern Processing Laboratory, UoA, in April 2021 under the supervision of Prof. Dr. Jungpil Shin. His research interests include computer vision, pattern recognition, and deep learning. He is currently working on human activity recognition, human gesture recognition, Parkinson’s disease diagnosis,  and ADHD diagnosis.

\end{IEEEbiography}

\begin{IEEEbiography}[{\includegraphics[width=1in,height=1.25in,clip,keepaspectratio]{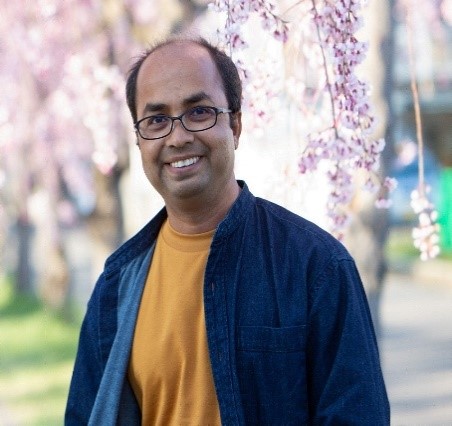}}]{MD. AL MEHEDI HASAN} (Member, IEEE) received the B.Sc., M.Sc., and Ph.D. degrees in computer science and engineering from the Department of Computer Science and Engineering, University of Rajshahi, Rajshahi, Bangladesh. He was a Lecturer, an Assistant Professor, an Associate Professor, and a Professor with the Department of Computer Science and Engineering, Rajshahi University of Engineering and Technology (RUET), Rajshahi. Recently, he completed a postdoctoral research with the School of Computer Science and Engineering, The University of Aizu, Aizuwakamatsu, Japan. His research interests include machine learning, deep learning, bioinformatics, health informatics, computer vision, probabilistic and statistical inference, medical image processing, sensor-based data analysis, human-computer interaction, operating systems, computer networks, and security.
\end{IEEEbiography}

\begin{IEEEbiography}[{\includegraphics[width=1in,height=1.25in,clip,keepaspectratio]{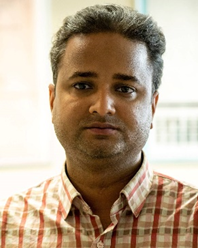}}]{MD. MANIRUZZAMAN} received the B.Sc., M.Sc., and M.Phil. degrees in statistics from the Department of Statistics, University of Rajshahi, Rajshahi, Bangladesh, in 2013, 2014, and 2021, respectively, and the Ph.D. degree in computer science and engineering from The University of Aizu, Japan, in 2024, under a scholarship from the Japanese Government (MEXT). He became a Lecturer and an Assistant Professor with the Statistics Discipline, Khulna University, Khulna, Bangladesh, in September 2018 and September 2020, respectively. He has co-authored more than 50 papers published in widely cited journals and conferences. His research interests include bioinformatics, artificial intelligence, pattern recognition, medical images, signal processing, biomedical data science, brain function, development disorders, machine learning, data mining, and big data analysis.
\end{IEEEbiography}

\EOD
\end{document}